\documentclass[smallcondensed]{svjour3}     
\smartqed  
\usepackage[utf8]{inputenc}
\usepackage{cite}
\usepackage{algorithm}
\usepackage{algpseudocode}
\usepackage[cmex10]{amsmath}
\usepackage{graphicx}
\usepackage{amsfonts, amssymb}
\usepackage{subfig}
\usepackage{float}
\usepackage[table,xcdraw]{xcolor}
\usepackage{adjustbox}
\usepackage{framed}
                     
\usepackage{multirow}
\usepackage{array}
\usepackage{makecell}
\usepackage{booktabs}
\newcolumntype{K}[1]{>{\centering\arraybackslash}p{#1}}
\begin{document}

\title{Test Case Prioritization Techniques for Model-Based Testing: A Replicated Study}



\author{João Felipe S. Ouriques \and Emanuela G. Cartaxo \and Patrícia D. L. Machado
}

\authorrunning{Ouriques et al.} 

\institute{J. F. S. Ouriques \and E. G. Cartaxo \and P. D. L. Machado \at Federal University of Campina Grande
              882, Aprígio Veloso \\
              Campina Grande, Brazil\\
              \email{jfelipe@copin.ufcg.edu.br}
           \and
           E. G. Cartaxo \at
	           \email{emanuela@copin.ufcg.edu.br}
           \and
           P. D. L. Machado \at
	           \email{patricia@computacao.ufcg.edu.br}
}

\date{Received: date / Accepted: date}

\maketitle


\begin{abstract}
	Recently, several Test Case Prioritization (TCP) techniques have been proposed to order test cases for achieving a goal during test execution, particularly, revealing faults sooner. In the Model-Based Testing (MBT) context, such techniques are usually based on heuristics related to structural elements of the model and derived test cases. In this sense, techniques' performance may vary due to a number of factors. While empirical studies comparing the performance of TCP techniques have already been presented in literature, there is still little knowledge, particularly in the MBT context, about which factors may influence the outcomes suggested by a TCP technique. 
	In a previous family of empirical studies focusing on labeled transition systems, we identified that the model layout, i.e. amount of branches, joins, and loops in the model, alone may have little influence on the performance of TCP techniques investigated, whereas characteristics of test cases that actually fail definitely influences their performance. However, we considered only synthetic artifacts in the study, which reduced the ability of representing properly the reality.
	In this paper, we present a replication of one of these studies, now with a larger and more representative selection of techniques and considering test suites from industrial applications as experimental objects. Our objective is to find out whether the results remain while increasing the validity in comparison to the original study.
	Results reinforce that there is no best performer among the investigated techniques and characteristics of test cases that fail represent an important factor, although adaptive random based techniques are less affected by it.
	
\end{abstract}

\section{Introduction}
\label{sec:introduction}

Software testing is an important activity to assess software quality. When performing this activity,
the involved personnel produces and maintains several artifacts,
such as test cases and fault reports. 
As a result, the testing environment and tasks usually demand a high amount of resources \cite{sommerville1}, which is particularly critical in industry. 
In this sense, 
much research has been carried out aiming at reducing the costs of software testing. Model-Based Testing (MBT) has been proposed as a way to achieve this goal.

MBT is the automation of the black-box testing design. Its main idea is to save time by generating test cases automatically from a behavioral model of the System Under Test (SUT). 
Since there is usually a number of possible ways to exercise a system,
 test case generation algorithms
can produce test suites whose execution costs can be prohibitive \cite{utting1}. On one hand, this problem is not particular to the MBT field. In practice, test suites may grow considerably in size and complexity, for instance during regression testing, as new functionalities are integrated into a system. On the other hand, MBT test suites can be impractical from inception. 

To address this problem, 
test case selection (TCS), test suite reduction (TSR) and test case prioritization (TCP) techniques have been proposed in the literature \cite{CatalM12}. 
TCS techniques aim at selecting a subset of the test cases for execution according to a specific goal, such as testing a modification performed in a system. On the other hand, TSR techniques focus on removing from the test suite test cases that are redundant w.r.t. a set of test requirements as long as the reduced test suite covers that requirements as minimally as possible. 
While the goal of 
TSR is to produce a more cost effective test suite, studies presented in the literature argue that TSR techniques may not work effectively, since they discard test cases and, consequently, some failures may not be revealed \cite{jeffrey1}. 

TCP techniques have been investigated in order to define an execution order for the test cases of a test suite according to a given testing goal, such as revealing faults as early as possible \cite{RUCH99}. Since TCP techniques do not discard test cases, the potential of the test suite to reveal faults is not decreased.
 These techniques are suitable for general development contexts, very related to situations that many (or all) test cases are new, and also more specific ones, such as regression testing, depending on the information taken into account by the techniques \cite{Rothermel01}. For regression testing, extensive research on TCP techniques, particularly based on the use of metaheuristics such as genetic algorithms have been proposed and evaluated in empirical studies \cite{HuangHCC10, RajuU12}. In such a context, test cases have already been executed in previous versions of the software and techniques make use of historical information such as test cases that fail to perform prioritization. However, an interesting finding about TCP techniques indicates that in the presence of new test cases, these techniques may behave differently \cite{LuLCZHZZ16}, so further investigation is needed to better understanding this context. Besides not all regression testing techniques are applicable to the general context, as their metaheuristics may depend on historical information.

Both code-based and model-based test suites may be handled by TCP techniques, although most techniques presented in the literature have been defined and evaluated for code-based suites in the context of regression testing \cite{elbaum3,jiang1}. On the other hand, TCP in the MBT context, particularly for general development settings, still deserves further investigation \cite{CatalM12}. Only few techniques and preliminary studies have been presented, making it difficult to assess current limitations of the techniques. 


To provide useful information for the development of new TCP techniques and to offer decision making evidence for practitioners, empirical studies should focus on controlling and/or observing factors that may determine the success of a given technique. Considering the goals of model-based TCP, a number of factors can be determinant such as the size and test suite coverage, the model's layout (that may determine the size and characteristics of test cases), the amount, distribution and characteristic of the test cases that fail, and the test cases degree of redundancy. 

In a previous empirical investigation, focusing on Labeled Transition Systems \cite{OuriquesJSERD}, we evaluated whether the \textit{amount of test cases that fail}, the \textit{model layout} (represented by the amount of branches, joins, and loops) and \textit{characteristics of test cases that fail} (if they traverse many or few branches, joins, or loops) affect the ability of revealing faults -- measured by the Average Percentage of Fault Detection (APFD) -- of a set of TCP techniques. We set up three different empirical studies to evaluate these factors and we found out that the model layout does not affect significantly the techniques. However, the characteristics of the test cases that fail, specifically the amount of branches/steps they exercise, affect the investigated techniques. 
The series of studies considered synthetic models to represent exactly the specific investigated conditions. However, even though we had observed the layout of industrial models (amount of transitions, states, branches, joins, and loops) to generate the investigated ones, they may not represent the reality and therefore, they would be considered as a threat to validity.

Wohlin et al. discuss the role of the empirical research on Software Engineering and propose an interesting framework for experiment conduction \cite{wohlin1}. The empirical evidence itself is already important, but we as scientists need to make sure that any other researcher is able repeat the same conditions and achieve similar results, increasing the evidence power or eventually rejecting it. Unfortunately most SE experiments have not been replicated \cite{GomezJV14}. Therefore, we need to perform replications in order to consolidate a body of knowledge build upon empirical evidence.

In this paper, \textbf{we replicate the third experiment from \cite{OuriquesJSERD}, comparing a set of general TCP techniques, investigating the influence of characteristics, particularly the size - or the amount of steps, of the test cases that fail}. The objective is to repeat the conditions evaluated in the original study, but involving more techniques and having industrial artifacts as objects, to verify if the results are repeatable and not artifactual (caused by the experimental artifacts). Likewise the original study, and as done by Henard et al. \cite{HenardPHJL16}, this replication study involves techniques that follow the simplest operation, which is adding one test case at a time in the prioritized test sequence, and do not use data from previous test executions (regression test prioritization techniques are out of our scope). All techniques in this study take the same artifact as input, which is a test suite.

The industrial applications considered in the study are modeled as use cases from which system test cases are automatically generated by the development team for manual testing execution. 
Despite the fact that a number of tools/techniques have already been developed to support test case automation, manual testing is still a relevant practice in industry, particularly, at system testing level \cite{HemmatiFM15}. Therefore, it is important to apply TCP (and other strategies) in the aforementioned context in order to deal with the involved costs. 
	
The results discussed in this paper point to the same direction of the ones already presented in the original investigation. Depending on the investigated characteristic of the test cases that fail, one technique performs better than the other ones, indicating some influence of it. Moreover, the study confirms that techniques based on random choices appear to be less affected by these different characteristics, although being less accurate. Besides, none of the investigated techniques is prevalent over the other ones with respect to the ability of unveil faults.
	
The remainder of the paper is structured as follows.
Section~\ref{sec:background} introduces fundamental concepts, along with the formal definition of the TCP problem. 
Section~\ref{sec:example} presents an example illustrating the problem that we are investigating. 
Section~\ref{sec:relatedWork} summarizes the research involving TCP techniques and empirical studies evaluating them.
Section~\ref{sec:techniques} shows an overview of the techniques investigated in this paper.
Section~\ref{sec:empirical} presents important details about the original study, as well as describes the replication we perform, detailing similarities and differences between the studies
and Section~\ref{sec:finalRemarks} presents final remarks about the results obtained and pointers for further research.
\section{Background}
\label{sec:background}

In this section we present an overview of the MBT process, including details about what kind of models we work in this investigation, and the TCP problem as well as its contexts of application.

\subsection{Model-Based Testing}
\label{sec:MBT}

\begin{figure}
	\centering
	\includegraphics[width=0.8\linewidth]{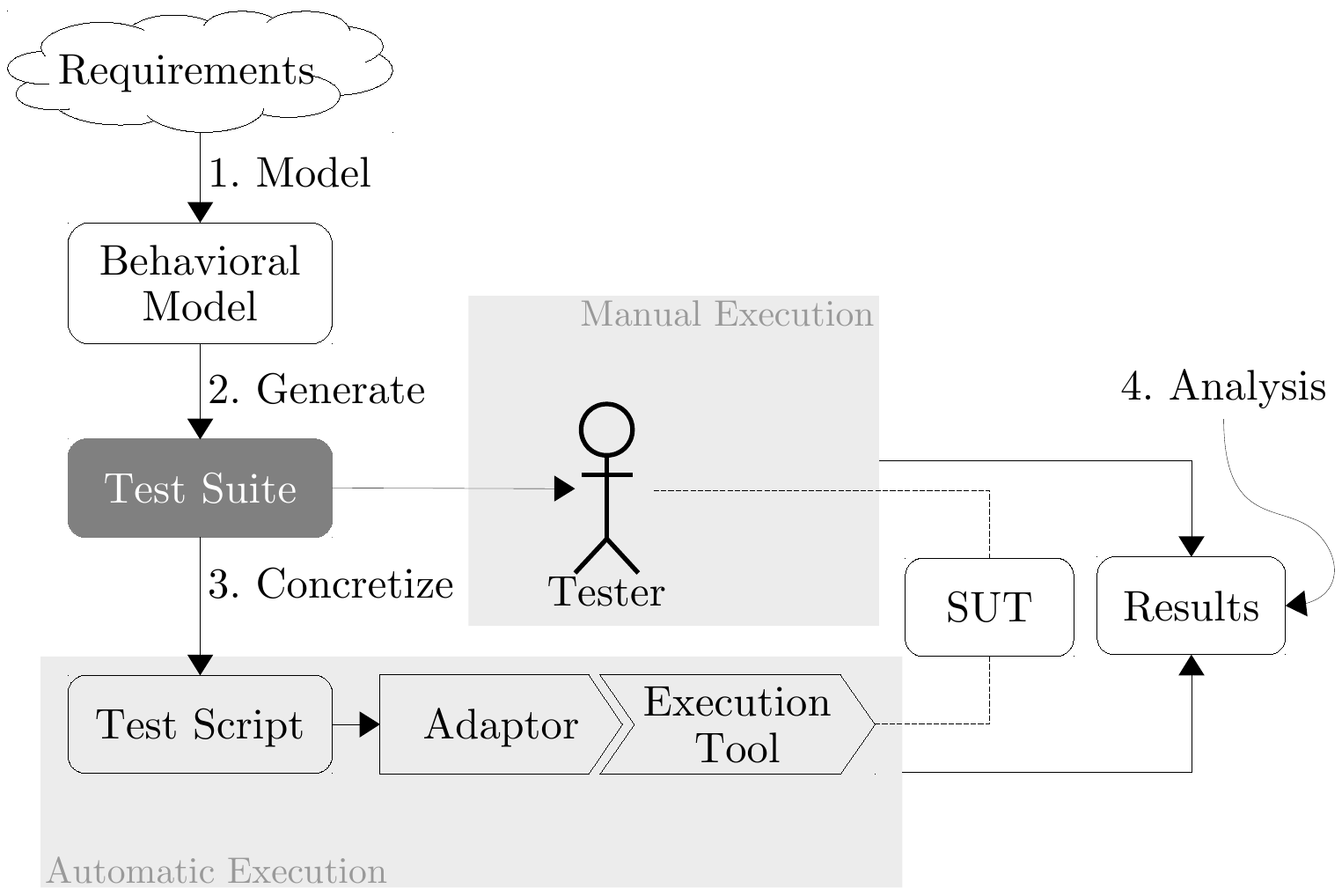}
	\caption{Activities (numbered labels) and artifacts (general shapes) involved in MBT.}
	\label{fig:MBT}
\end{figure}

MBT is the automation of the design of black-box testing and it is largely used in the \textbf{System Testing level} \cite{utting1}. 
From Figure~\ref{fig:MBT}, the process 
takes as input the requirements that the system must satisfy. The first step is \textbf{modeling} these requirements
(Section~\ref{sec:modeling}). Then, in the next step, one can \textbf{generate} test suites automatically by exploring the behavioral model 
(Section~\ref{sec:generation}). While generated test suites may be ready for manual execution, 
in the \textbf{concretize} step, test suites may be refined with more details to be executed automatically, e.g. extra code connecting test cases to the SUT, more complex data and verdicts.  Both kinds of test execution produce reports containing the reported faults and the test cases that fail because of them, i.e. test cases that reveal these faults, as result and regardless the kind of execution, testers still need to \textbf{analyze} these results, e.g. which portions of the SUT need debugging or refactoring for correcting faults.
%
%
In the experiments presented in this paper, 
test suites for manual execution
are the main input. 

\subsubsection{Modeling Systems}
\label{sec:modeling}

In MBT, the main artifact is the system model representing its behavior through its execution flows. Based on it, the subsequent testing activities, for example test case generation, take place. The first decision is to define the model notation, based on the abstraction and the testing level. In this work, since we consider a high abstraction level in system testing, a Transition-Based 
notation is 
used because of its visual representation, availability of support tools and 
adequacy for abstract control-flow representation \cite{utting1}. More specifically, we use Labeled Transition Systems (LTS) as our model representation. 
Formally, an LTS is a 4-tuple $<S, L, T, s_0>$, where \cite{VriesT00}:

\begin{itemize}
	\item $S$ is a non-empty and finite set of \textit{states};
	\item $L$ is a finite set of \textit{labels};
	\item $T \subseteq SxLxS$ is a set of triples, the \textit{transition relation};
	\item $s_0$ is the \textit{initial state}. 
\end{itemize}

\begin{figure}
	\centering
	\includegraphics[width=0.8\linewidth]{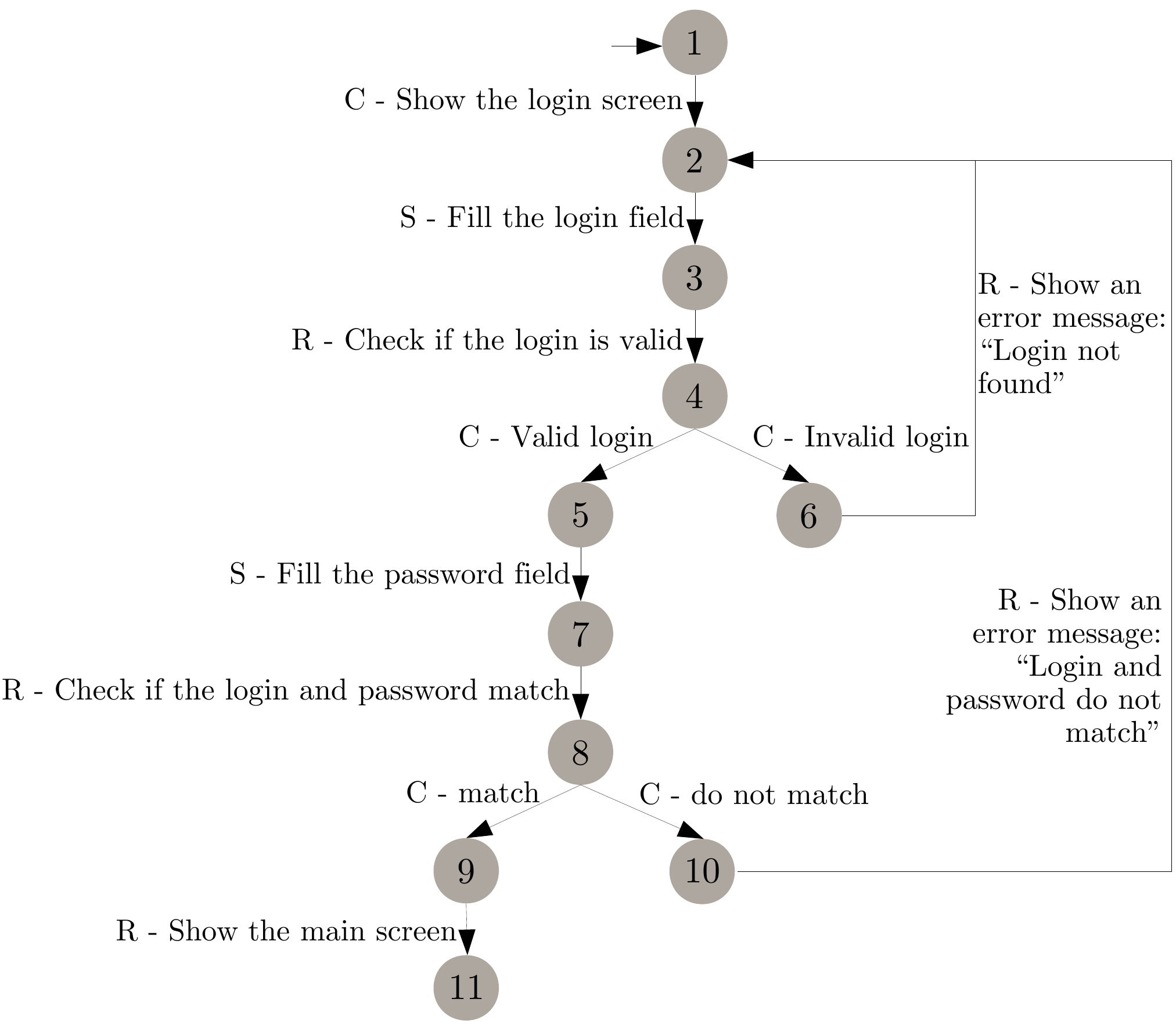}
	\caption{LTS representing a use case for login and password verification.}
	\label{fig:ExampleModel}
\end{figure}

To model a system using this notation, \textbf{user steps}, \textbf{system responses}, and  \textbf{system conditions} are represented
as transition's labels. To represent a condition or decision, 
it is necessary to branch the state and add a transition representing each alternative. For the sake of notation, the transition label has  
reserved prefixes: i) \textit{``S - "} for user steps; ii) \textit{``R - "} for system responses;  and iii) \textit{``C - "} for conditions. It is also possible to represent the junction of more than one flow and loops. 
In the context of our work, we use LTS to model use cases. A use case may present three kinds of flows:
i) \textit{base}, describing the most common use scenario that is completed successfully;
ii) \textit{alternative}, defining a user's alternative behavior or a different way that the user has to do something; and
iii)\textit{exception}, specifying the occurrence of an error returned by the system. In Figure~\ref{fig:ExampleModel}, one can see an example of a use case that verifies login and password modeled as an LTS, comprising one base and two exception flows.

\subsubsection{Test Case Generation}
\label{sec:generation}

One of the most important aspects about MBT is the automation of the test suite generation activity \cite{utting1}. To do so, algorithms usually explore the application models to produce sequences of tasks to be performed during test execution. 

In 
the context of our work, we consider a depth first search (DFS) algorithm 
that traverses an LTS recording the paths from the initial state -- each path recorded is a test case. 
For instance, according to the \textbf{all-n-loop-paths} criterion \cite{CoutinhoCM14},
the algorithm decides the paths' end, by traversing
the LTS until a given loop is recorded a specific amount of times. 
Since the LTS models a use case, the test case presents pairs of steps and system responses, as well as conditions that the system must satisfy, providing the tester the ability to compare the system outputs to the responses expressed by the model as well as determining the conditions under which the test case should be executed. 
The output of this step is a test suite that can be manually executed but, because of the exhaustive nature of the generation algorithm, its complete execution may not be appropriate.

\subsection{Test Case Prioritization}
 
TCP techniques reorder test cases of a test suite to achieve an objective as soon as possible in the testing process. Formally, the TCP problem is defined as follows \cite{elbaum3}:

\begin{center}
\textit{Let} $TS$ be a test suite, $PTS$ be a set of every permutation of $TS$ and $f : PTS \rightarrow \mathbb{R}$.
\textit{Find a} $TS'$ $\in$ $PTS$ $\mid$
$\forall$ $TS''$ ($TS''$  $\in$ $PTS$)
($TS''$ $\neq$ $TS'$) $\cdot$ $f$($TS'$)
$\geq$ $f$($TS''$).
\end{center}

\noindent where $f$ is a function that measures whether a prioritized test suite achieves the prioritization objective. TCP algorithms aim at defining a prioritization order that maximizes this function.


TCP has two clear challenges: 
i) analysing the whole set of all permutations of $TS$; and ii) defining the evaluation function $f$. Since the number of permutations of a set $A$ is $|A|!$, analyzing the elements from $PTS$ might be unfeasible if $TS$ is big, as discussed by Lima et al. \cite{LimaISA2009}. Thereby, usually TCP techniques aim at selecting one test case at a time, placing it in the desired order according to certain heuristics. Moreover, the evaluation function is related to the prioritization goal, for example, to reduce test case setup time \cite{LimaISA2009} and, more commonly, to accelerate fault detection. Depending on the goal, the information to define $f$ is not available beforehand, making its precise definition impossible. For instance, considering fault detection, information regarding faults and/or which test cases detect them is not available in advance, then a technique is not able to maximize it. Therefore, techniques that focus on fault detection, estimate fault information through surrogates \cite{YooHTS2009}.

TCP is suitable for code-based and model-based contexts, but it has been more applied in the code-based context, related more often to regression testing. In this sense, Rothermel et al. \cite{Rothermel01} proposed the following classification:
	
\begin{itemize}
	\item General test case prioritization - TCP is applied any time in the software development process, even in initial testing activities;
	\item Regression testing prioritization - TCP takes place after performing a set of changes in the SUT.
	Therefore, TCP techniques can use information gathered from previous runs of existing test cases to leverage the prioritization of test cases for subsequent runs.
\end{itemize}
	
In this paper, \textbf{we focus on system testing level, model-based general test case prioritization}. Therefore, we are not considering that any other information is available, besides the test suites. Regression testing approaches may also consider test case execution history and/or applied modifications.

\subsection{Replication Studies}
\label{sec:replicationStudies}

The act of repeating a study is an important part of the scientific method \cite{JuristoM2001, wohlin1}. Although there is no consensus about definitions, a replication study consists of repeating as closely as possible the conditions observed in the original one. Furthermore, replications are often used as a method to assess whether a set of findings is stable enough to be discovered more than once \cite{GomesJV10}.

In a further investigation, Gómez et al. \cite{GomezJV14} have reviewed several ways of classifying replication studies, considering software engineering and other disciplines. Based on the discussed aspects, they suggest a new classification based on four dimensions. Ahead, we provide a general understanding about them, along with a brief way of approaching them by comparing the original and replication study: 

\begin{itemize}
	\item \textbf{Operationalization}: this dimension addresses the way that constructs are translated into their manifestations in the replication, i.e. how the cause/effect relation is 
	translated into the studies?;
	\item \textbf{Population}: it details aspects of subjects and objects, i.e. which aspects of the subjects and experimental objects are 
	investigated in the studies?; 
	\item \textbf{Protocol}: 
	encompasses materials, experimental objects, forms, and procedures for the replication, i.e. are the studies considering the same metrics? How about the experiment design? And data analysis techniques?;
	\item \textbf{Experimenters}: regarding the people involved in the experiment, i.e. is the same team responsible for both studies? Is the study somehow dependent on the involved researchers?
\end{itemize}

Besides, the authors classify replications in types, based on the dimensions already discussed. Therefore, we classify our replication using these dimensions in Section~\ref{sec:replicationStudy}.

\section{Motivating Example}
\label{sec:example}

The effectiveness of a TCP technique is often evaluated by the ability of revealing failures as soon as possible in the testing activity. Ideally, a technique in this scenario should put test cases that fail in the first positions of the prioritized test suite.

Disregarding whatever previous knowledge about failures of the system may be available, which is out of the scope of this paper, most TCP techniques for general prioritization are based on structural aspects and also make some assumptions for ordering the test cases. For example, the longer the test case is or the more branches it traverses, the higher is the probability of revealing failures.

To illustrate how TCP techniques may be affected by factors related to characteristics of test cases that fail, consider the test suite presented in Table~\ref{tab:testcasesexample}. This test suite is generated from the model in Figure~\ref{fig:ExampleModel}, by the action of a generation algorithm instrumented with a coverage criterion that traverses a loop at most twice. Moreover, consider that when the test cases are executed, only the scenario that represents the successful login fails, in other words, TC1 fails.

\begin{table}[htp]
\scriptsize
\centering
\caption{\textbf{Test cases generated from the model in Figure~\ref{fig:ExampleModel}.}}
\label{tab:testcasesexample}
\begin{tabular*}{\textwidth}{|l|p{270pt}|c|}
	\hline
	\textbf{Label} & \textbf{Steps} & \textbf{Size}\\
	\hline
	TC1 &
	C - Show the login/password screen
	$\rightarrow$ S - Fill the login field
	$\rightarrow$ R - Check if the login is valid
	$\rightarrow$ C - Valid login
	$\rightarrow$ S - Fill the password field
	$\rightarrow$ R - Check if the login and password match
	$\rightarrow$ C - match
	$\rightarrow$ R - Show the main screen of the application  & 8\\
	\hline
	TC2 &
	C - Show the login/password screen
	$\rightarrow$ S - Fill the login field
	$\rightarrow$ R - Check if the login is valid
	$\rightarrow$ C - Valid login
	$\rightarrow$ S - Fill the password field
	$\rightarrow$ R - Check if the login and password match
	$\rightarrow$ C - do not match
	$\rightarrow$ R - Show error message: ``Login and password do not match"
	$\rightarrow$ S - Fill the login field
	$\rightarrow$ R - Check if the login is valid
	$\rightarrow$ C - Valid login
	$\rightarrow$ S - Fill the password field
	$\rightarrow$ R - Check if the login and password match
	$\rightarrow$ C - match
	$\rightarrow$ R - Show the main screen of the application & 15\\
	\hline
	TC3 &
	C - Show the login/password screen
	$\rightarrow$ S - Fill the login field
	$\rightarrow$ R - Check if the login is valid
	$\rightarrow$ C - Valid login
	$\rightarrow$ S - Fill the password field
	$\rightarrow$ R - Check if the login and password match
	$\rightarrow$ C - do not match
	$\rightarrow$ R - Show error message: ``Login and password do not match"
	$\rightarrow$ S - Fill the login field
	$\rightarrow$ R - Check if the login is valid
	$\rightarrow$ C - Valid login
	$\rightarrow$ S - Fill the password field
	$\rightarrow$ R - Check if the login and password match
	$\rightarrow$ C - do not match
	$\rightarrow$ R - Show error message: ``Login and password do not match" & 15\\
	\hline
	TC4 &
	C - Show the login/password screen
	$\rightarrow$ S - Fill the login field
	$\rightarrow$ R - Check if the login is valid
	$\rightarrow$ C - Valid login
	$\rightarrow$ S - Fill the password field
	$\rightarrow$ R - Check if the login and password match
	$\rightarrow$ C - do not match
	$\rightarrow$ R - Show error message: ``Login and password do not match"
	$\rightarrow$ S - Fill the login field
	$\rightarrow$ R - Check if the login is valid
	$\rightarrow$ C - Invalid login
	$\rightarrow$ R - Show error message: ``Login not found" & 12\\
	\hline
	TC5 &
	C - Show the login/password screen
	$\rightarrow$ S - Fill the login field
	$\rightarrow$ R - Check if the login is valid
	$\rightarrow$ C - Invalid login
	$\rightarrow$ R - Show error message: ``Login not found"
	$\rightarrow$ S - Fill the login field
	$\rightarrow$ R - Check if the login is valid
	$\rightarrow$ C - Valid login
	$\rightarrow$ S - Fill the password field
	$\rightarrow$ R - Check if the login and password match
	$\rightarrow$ C - match
	$\rightarrow$ R - Show the main screen of the application & 12\\
	\hline
	TC6 &
	C - Show the login/password screen
	$\rightarrow$ S - Fill the login field
	$\rightarrow$ R - Check if the login is valid
	$\rightarrow$ C - Invalid login
	$\rightarrow$ R - Show error message: ``Login not found"
	$\rightarrow$ S - Fill the login field
	$\rightarrow$ R - Check if the login is valid
	$\rightarrow$ C - Valid login
	$\rightarrow$ S - Fill the password field
	$\rightarrow$ R - Check if the login and password match
	$\rightarrow$ C - do not match
	$\rightarrow$ R - Show error message: ``Login and password do not match" & 12\\
	\hline
	TC7 &
	C - Show the login/password screen
	$\rightarrow$ S - Fill the login field
	$\rightarrow$ R - Check if the login is valid
	$\rightarrow$ C - Invalid login
	$\rightarrow$ R - Show error message: ``Login not found"
	$\rightarrow$ S - Fill the login field
	$\rightarrow$ R - Check if the login is valid
	$\rightarrow$ C - Invalid login
	$\rightarrow$ R - Show error message: ``Login not found" & 9\\
	\hline
\end{tabular*}
\end{table}

Suppose the application of two basic prioritization techniques, considering test case size (the column Size in Table~\ref{tab:testcasesexample}) as the test requirement and fault detection as the test goal: 1) \textbf{greedy}, which chooses iteratively test cases with the highest amount of steps and 2) \textbf{reverse-greedy}, which chooses iteratively test cases with the lowest amount of steps. By applying these two techniques, we can obtain the orders presented in Table~\ref{tab:exampleresults}. Note that \textbf{greedy} places the failure represented by TC1 near to the end of the sequence -- a poor result for a TCP technique. On the other hand, \textbf{reverse-greedy} places the failure in the first position -- conversely, a desirable behavior for a prioritization technique. TC1 has the fewest number of steps along with TC7.

\begin{table}
\scriptsize
\centering
\caption{Test cases order produced by the greedy and reverse-greedy techniques for the test suite from Table~\ref{tab:testcasesexample}.}
\label{tab:exampleresults}
\begin{tabular}{|c|c|}
	\hline
	\textbf{Technique} & \textbf{Test cases sequence}\\
	\hline
	Greedy & TC2, TC4, TC6, TC3, TC5, \textbf{TC1}, TC7\\
	\hline
	Reverse-Greedy & \textbf{TC1}, TC7, TC4, TC7, TC3, TC6, TC2\\
	\hline
\end{tabular}
\end{table}

In this example, it is easy to identify the factor that has influenced the results. However, for more sophisticated techniques and more complex test suites, it is not always obvious to identify all factors involved and how each particularly technique can be affected by them.

To make TCP techniques applicable in practice, it is important to understand why one is more successful than other. Due to the amount of steps in the test case that fails? Or are there more factors that have influence on the results? We began to investigate these questions and we found out that the model layout is not much determinant for the technique's success. On the other hand, characteristics of test cases that fail can have a considerable influence. Since we investigated only synthetic models, test suites, and faults in our previous studies \cite{OuriquesJSERD}, in this work we extend one of the studies by considering more techniques and applying a more realistic approach with industrial applications and faults.
\section{Related Work}
\label{sec:relatedWork}

This paper presents a replicated study of the third experiment from \cite{OuriquesJSERD}, comparing a set of general TCP techniques in the context of general prioritization focusing on MBT test suites. A detailed comparison between the studies from both papers are presented in Section~\ref{sec:empirical}. In this section, we review other related work regarding TCP techniques and empirical studies presented in the literature.

Most of TCP techniques already proposed focus on code-based test suites and the regression testing context \cite{ElbaumRKM04,KTH05,CatalM12}. Besides, the experimental studies 
presented so far have discussed whether a technique is more effective than others, comparing them, for example by their ability to reveal faults through the Averaged Percentage of Fault Detection - APFD - metric. Concerning these studies, there is no indication of general results, which evidences the need for further investigation and empirical studies that may contribute to advances in the state-of-the-art.
	
Regarding code-based prioritization, Zhou et al. \cite{ZhouSS12} compare failure-detection capabilities of the Jaccard-distance-based ART and Manhattan-distance-based ART. The authors use branch coverage information and the results showed that, for code-based test suites, Manhattan distance is more effective than Jaccard \cite{ZhouSS12}. Jeffrey and Gupta \cite{jeffrey1} propose an algorithm that prioritizes test cases based on coverage of statements in relevant slices and discuss insights from an experimental study that considers also total coverage. Furthermore, Henard et al. \cite{HenardPHJL16} perform a study comparing white-box and black-box TCP techniques. They claim that diversity-based techniques perform best among black-box ones, and, even though white-box techniques have more information about the SUT available, they do not present a significantly higher performance than black-box ones. This result suggests the need for further investigation about the behavior of black-box MBT techniques.

Moreover, Do et al. \cite{Do10} present a series of controlled experiments evaluating the effects of time constraints and faultiness levels on the costs and benefits of TCP techniques. They define faultiness level as a variable that manipulates the amount of faults (mutants) randomly placed in applications. The results show that time constraints can significantly influence both the cost and effectiveness. Moreover, when there are time constraints, the effects of increased faultiness are stronger.

Furthermore, Elbaum et al. \cite{elbaum3} compare the performance of five prioritization techniques in terms of effectiveness and show how the results of this comparison can be used to select a technique (regression testing) \cite{ElbaumRKM04}. They compare techniques that take into account the coverage of functions in the source code (total and additional coverage), modifications between two versions, and feedback of functions already covered as guide to prioritize test cases. They apply these techniques to eight programs and their characteristics (such as number of versions, KLOC, number and size of the test suites, and average number of faults) are taken into account. Hao et al. \cite{HaoZZWWX16} also provide evidence favoring the technique based on the additional coverage of code elements, arguing that covering optimally code elements does not lead to relevant gains compared to the additional coverage.

By considering the use of models in the regression testing context, Korel et al. \cite{KTH05, Korel07, KKT08} present two model-based TCP methods: selective test prioritization and model dependence-based test prioritization. Both techniques focus on modifications made to the system and models. The inputs are the base version of the system model, modeled through an Extended Finite State Machine, and the delta version. On the other hand, our focus is on general TCP techniques, as defined by Rothermel et al. \cite{Rothermel01}, where modifications are not considered.

Generally, in the MBT context, one can find proposals to apply general TCP from UML activity diagrams: Sapna and Mohanty \cite{SapnaM09}, Kundu et al. \cite{KunduSSM09}, and Kaur et al. \cite{KaurBS12}. Since these techniques are based on a simple strategy of placing one test case in the prioritized sequence at a time, we investigated them in our study, as well as other code-based techniques with loose requirements on code elements.

Other authors investigate the use of soft-computing methods to solve the TCP problem. Fevzi Belli, Mubariz Eminov, and Nida Gokce present a long term research in this context \cite{GokceEB2006, BelliEG2007, BelliG2010}. They propose techniques that prioritize test cases by clustering model elements (events from Event Sequence Graphs) and giving preference degrees for the test cases based on the importance of the events they cover: the first one using an unsupervised neural network; the second one, a clusterer using a fuzzy version of the c-means algorithm; and the last one, the Gustafson-Kessel clustering algorithm. In their empirical investigation, they propose that the use of soft-computing might present a high computational cost, while none of them perform significantly better than the others. In the same context, several techniques based on genetic algorithms have been proposed \cite{HuangHCC10, RajuU12}. However, they also have historical artifacts as input to the fitness function. 

On the other hand, Sabharwal et al. \cite{SabharwalSS2010} and Nejad et al. \cite{NejadAD16} suggest techniques also based on genetic algorithms, but only based on static elements of test cases generated from UML activity diagrams (e.g. nodes and edges). Although they rely on similar input information, the underlying theory is different from our context of investigation. 
We intend to consider these techniques in a further study including techniques that use soft-computing methods.

There is a study very related to ours, in which the authors compare black-box system-level TCP techniques that operate over code-based test suites. Hemmati et al. \cite{HemmatiFM15} compare three techniques, each one representing a different approach: \textbf{topic coverage}, which prioritizes test cases to cover topics (linguistic analysis on the source code of the tests) as soon as possible; \textbf{text diversity}, which prioritizes test cases based on the distances between their string representations, using Euclidean and Manhattan functions to calculate these distances; \textbf{risk-driven}, which prioritizes test cases based on information about their previous executions. As a result, the authors suggest that none of the investigated approaches are highly dominant over the others. However, when historical information is available as in regression testing context, the risk-based approach is clearly superior. The authors also suggest repeating the study, or performing a similar one, on non-code-based test cases to increase external validity of the proposed results. Besides, Lu et al. \cite{LuLCZHZZ16} analyzed the behavior of several TCP techniques, including \textbf{Adaptive Random}, \textbf{Search-Based}, and \textbf{Coverage-Based}, during the evolution of eight java projects through a series of commits of their real development. Among their conclusions, they suggest that when new test cases are added in the test suite, the investigated techniques perform differently, which encourages us to follow investigating the scenario that no information about previous executions is available.

In summary, the contributions of this work are: i) to gather a set of TCP techniques based on selecting one test case at a time strategy, without historical information, in a common execution environment, 
focusing on model-based test suites and; ii) to revisit the empirical studies presented in \cite{OuriquesJSERD} and replicate one of them, considering test suites from industrial applications, exposing the influence of characteristics of the test cases that unveil faults on the TCP techniques.

\section{Techniques}
\label{sec:techniques}

In our research, we compare a set of techniques suitable to prioritize model-based test suites that do not resort to historical information, even though some of them underwent some adaptation to fit our research context. Each technique has a common input requirement (a test suite as input) and provides the same result (the prioritized test suite, according to their particular operation).  It is important to remark that some techniques investigated in this paper are part of a family, where members differ by a single configuration aspect or a distance function. Instead of considering just one representative of each family, we considered in the study all the variations described in this section in order to verify whether they have influence on the results, as done by \cite{CoutinhoCM14, Zhou10, LedruPBM12}. 

Based on this common ground, we investigate the following (family of) techniques.

\paragraph{\textbf{Optimal.}}
\label{tech:opt}

Empirical evaluations frequently include this technique as upper bound for the effectiveness of investigated techniques. It presents the best result that a technique is able to achieve with respect to the investigated evaluation metric. To obtain the best result, this technique must access key information required to maximize the performance, for example, if one performs an empirical evaluation with respect to fault detection, the optimal technique must take the untreated test suite and the failure report as inputs. In our study, we identify this technique as \textbf{Opt}. 



\paragraph{\textbf{Random.}}
\label{tech:ran}

This technique defines randomly the next test case to be placed in order until the untreated test suite is empty. Despite the fact that random choice can lead to optimal results by chance, experiments with TCP techniques have applied it as a lower bound control technique \cite{jiang1}. We identify it as \textbf{Ran} in our investigation. 


\paragraph{\textbf{Adaptive Random Prioritization - ARP.}}
\label{tech:arp}

This family of techniques prioritizes the input test suite by placing one test case in order at a time, using a notion of distance to spread more evenly the test cases \cite{jiang1, Zhou10}. To do so, it manipulates two structures: the prioritized sequence and the candidate set. ARP has a representative randomness in the generation of this candidate set, since the next test case to be placed in order comes from the candidates. Therefore, even though it applies the notion of distance, the random candidate selection still plays an important role on ARP.  

%


Jiang et al. \cite{jiang1} proposed ARP using 
\textit{jaccard} distance and 
the \textit{maximum of the minimum} distances between the candidates and the already prioritized test cases, as the notion of distance (we identify it in our study as \textbf{ARPJac}). 
In turn, Zhou \cite{Zhou10} suggested using 
\textit{manhattan} distance and keeping 
the same maximum of the minimum concept as the previous one (we label this variation as \textbf{ARPMan}). 


Besides, Coutinho et al. \cite{CoutinhoCM14} proposed a function to measure the similarity between model-based test cases. Then, we contribute to the investigation of ARP techniques by considering it 
as distance function, and considering 
the \textit{maximum of the minimum} similarities (\textbf{ARPSim1}), as well as the \textit{maximum of the maximum} similarities between the candidates and the already prioritized test cases (\textbf{ARPSim2}).



\paragraph{\textbf{Fixed Weights.}} 
\label{tech:fw}

Sapna and Mohanty \cite{SapnaM09} proposed a prioritization technique based on UML activity diagrams. It generates test cases from the input diagram and prioritizes these test cases by sorting them, using weights assigned to the input diagram elements. For the nodes, the technique assigns weight 3 for fork-join nodes, 2 for branch-merge nodes, 1 for action/activity nodes and for edges. 

Since in our context, we do not have the same elements that they considered from activity diagram, we adapt the technique to consider just the test suite, assigning 1 to simple steps and 2 to branch and join steps, and adding the weight of the step as calculated in the original technique (we label our adapted version in the study as \textbf{FW}). 

%

\paragraph{\textbf{STOOP.}}
\label{tech:stoop}

Kundu et al. \cite{KunduSSM09} proposed a technique called \textbf{Stoop}, also based on UML Sequence Diagrams. Likewise FixedWeights, STOOP is based on activity diagram elements, such as activity nodes and edges. This technique may be guided by three metrics for different prioritization objectives and, for our study we just consider the Averaged Weight Path Length - AWPL, since it is related to fault detection. AWPL is expressed $AWPL(tc)=\dfrac{\sum_{i=1}^{m}{eWeight(S_i)}}{m}$, where: $tc = (S_1, S_2, \cdots, S_m)$ is a test case with $m$ steps; and $eWeight(S_i)$ is a function that calculates the weight of the $i^{th}$ step, which is the amount of test cases that cover the given step.

We also had to adapt STOOP, since our context does not consider activity diagrams. The algorithm calculates AWPL for each test case and sorts them in descending order. We label our modified version as \textbf{Stoop} in the empirical investigation).


\paragraph{\textbf{Path Complexity.}}
\label{tech:pc}
Proposed by Kaur et al. \cite{KaurBS12}, this technique prioritizes test cases generated from UML Activity Diagrams. It calculates some properties for each test case, such as test case size and information flow metric, sums these properties to derive their complexities, and sorts the test cases decreasingly by complexity. 


This technique was also adapted since the investigated applications were not modeled through activity diagrams. We calculate the weight of the steps instead of the nodes and $fanin$ and $fanout$ represent the amount of steps executed respectively before and after the referred one, considering the whole test suite. In our study, we identify this technique as \textbf{PC}. 


\paragraph{\textbf{String Distance.}}
\label{tech:sd}

Ledru et al. \cite{LedruPBM12} proposed prioritizing test cases based on the resemblance between string representations of the involved test cases. It compares the inputs of test cases or even the body of JUnit test cases, since they might be faced as sequences of strings. 


The authors propose using four different functions: Hamming, Euclidian, Manhattan, and Levenshtein distances. Among these well-known functions, we consider all but Levenshtein, because it is very time consuming and leads to a result comparable to Manhattan distance \cite{LedruPBM12}. Therefore, we refer to these variations in our study as \textbf{SDh}, \textbf{SDe}, and \textbf{SDm}, respectively.

\paragraph{\textbf{Total and Additional Coverage of Steps.}}
\label{tech:stsa}

Elbaum et al. \cite{elbaum3} suggested using a greedy reasoning to prioritize test cases. The total approach sorts the test cases by the total amount of code statements that they cover, whereas the additional one also sorts them, but adjusts iteratively the statements already covered by the currently test case sequence, i.e. the next test case in order is the one that cover more statements not yet covered so far. Since our test suites are based on high abstraction steps instead of code statements, we adapted the techniques for our context by considering the coverage of steps. We identify these two techniques as \textbf{ST} and \textbf{SA} in our study.
\section{Empirical Investigation}
\label{sec:empirical}

In this section, we present the replication study.  For this, we follow guidelines suggested by Carver \cite{Carver10} to report replications. In Subsection \ref{sec:originalStudy}, we discuss important details about the original study, such as its variables, setup and results, whereas in Section \ref{sec:replicationStudy} we describe the replicated study. In Section \ref{sec:dataAnalysis}, we present the results obtained in the replication study, whereas in Section \ref{sec:resultsAndDiscussion}, we discuss these results and compare to the ones obtained in the original study. Furthermore, in Section \ref{sec:threatsToValidity}, we discuss threats to validity of the replication study.


\subsection{Background to this Replication - The Original Study}
\label{sec:originalStudy}

The original study was part of a family of studies that focused on exposing the influence of some factors on the ability of revealing faults of TCP techniques in the MBT context. More specifically, it investigated 
``how \textbf{general TCP techniques} behave when \textbf{test cases with certain properties fail}?". To investigate this question we needed to control the variables that represent the bold-faced aspects of the research question. Therefore, we considered the following variables and their possible values:

\paragraph{Independent Variables}

\begin{itemize}
	\item General prioritization techniques: \textbf{ARPJacMaxMin}, \textbf{ARPManMaxMin}, \textbf{FixedWeights}, and \textbf{Stoop};
	
	\item Characteristics of the test cases that fail;
	\begin{itemize}
		\item Longest test cases, i.e. the ones that comprise more transitions (\textbf{LongTC});
		\item Shortest test cases, i.e. the ones that comprise fewer transitions  (\textbf{ShortTC});
		\item Test cases that traverse more branches (\textbf{ManyBR});
		\item Test cases that traverse fewer branches (\textbf{FewBR});
		\item Test cases that traverse more joins (\textbf{ManyJOIN});
		\item Test cases that traverse fewer joins (\textbf{FewJOIN});
		\item Essential test cases, i.e. the ones that uniquely cover a specific transition in the model (\textbf{Essential});
	\end{itemize}

	\item Number of test cases that fail: fixed value equals to 1;
\end{itemize}

\paragraph{Dependent Variable}

\begin{itemize}
	\item Average Percentage of Fault Detection - $APFD=1-\frac{TF_1+TF_2+\ldots+TF_m}{nm}+\frac{1}{2n}$, where $TF_i$ is the position of the first test case that reveals the $i$-th fault, $m$ is the number of faults that the test suite is able to reveal, and $n$ is the size of the test suite \cite{elbaum3}.
\end{itemize}

\begin{figure}
	\centering
	\includegraphics[width=0.6\linewidth]{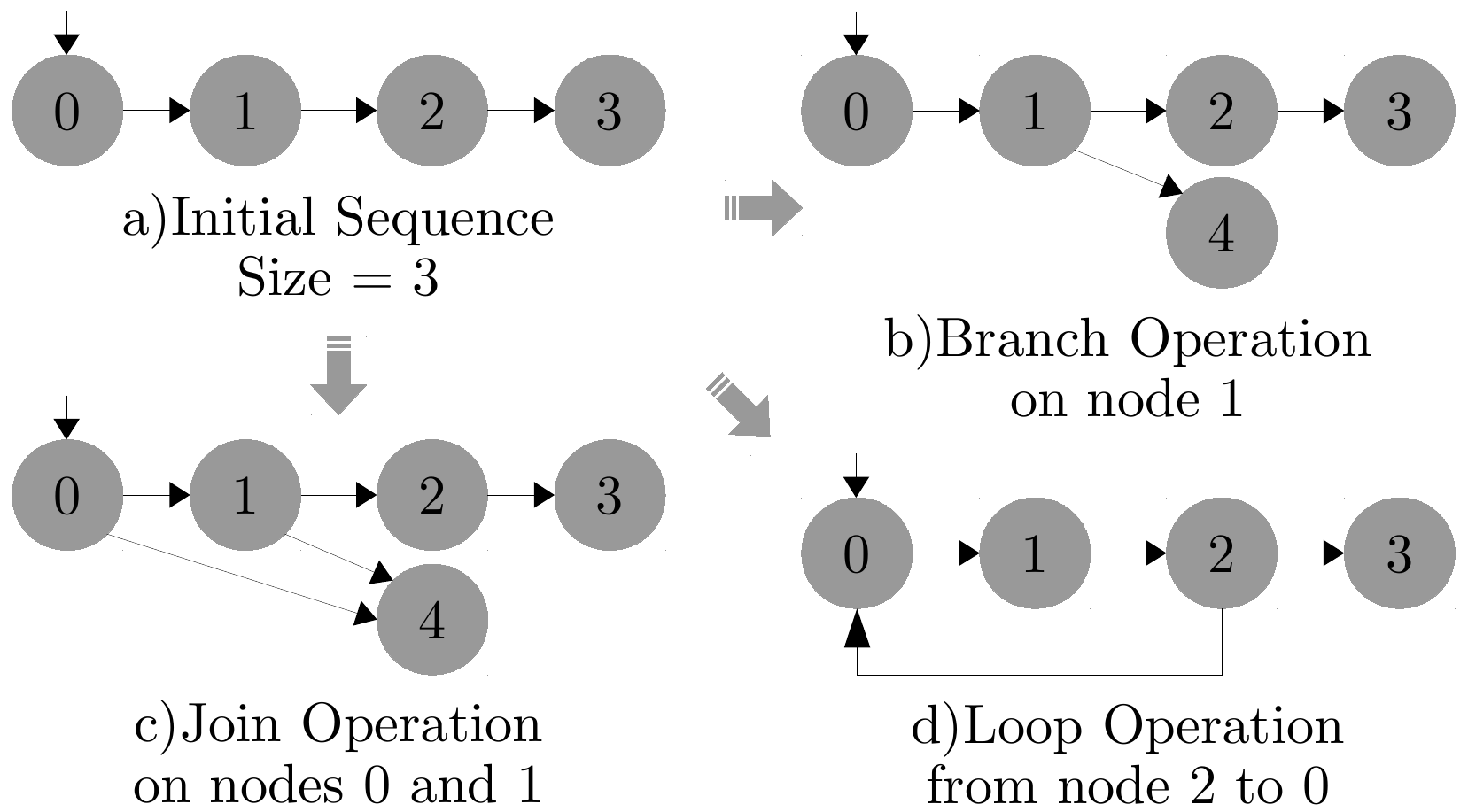}
	\caption{Operations performed by the LTS generator.}
	\label{fig:LTSGen}
\end{figure}

In MBT, test cases are generated from the system model as discussed in Section~\ref{sec:MBT}. In order to acquire a set of system models with similar characteristics for this study, keeping this influence as low as possible, we used a generator of synthetic LTS that 
can be configured by the number of branches, joins, and loops required. From an initial sequence of transitions, the generator performs the operations depicted in Figure~\ref{fig:LTSGen}, varying their order, to produce slightly different models.

Since we were not observing variations of system model itself, we considered a set of 31 artificial LTS as system models, generated using the same set of operations, which were 30 branches, 15 joins, 1 loop and maximum depth 25. We defined these values observing industrial applications models also modeled as LTS. Although the synthetic models did not represent the behavior of an actual system, they are structurally similar.  

Moreover, in order to control tightly the characteristics of the test cases that fail, we defined fault models according to the characteristic represented by the value of the variable. For instance, considering the \textbf{LongTC} characteristic, the algorithm sorted the test cases decreasingly by length (or number of steps). If there are more than one with the biggest length (same profile), one of them was chosen randomly. For example, if in a test suite the longest test cases has 15 steps, the algorithm selected randomly one of the test cases with size equals to 15.

\begin{figure}
	\centering
	\includegraphics[width=0.8\linewidth]{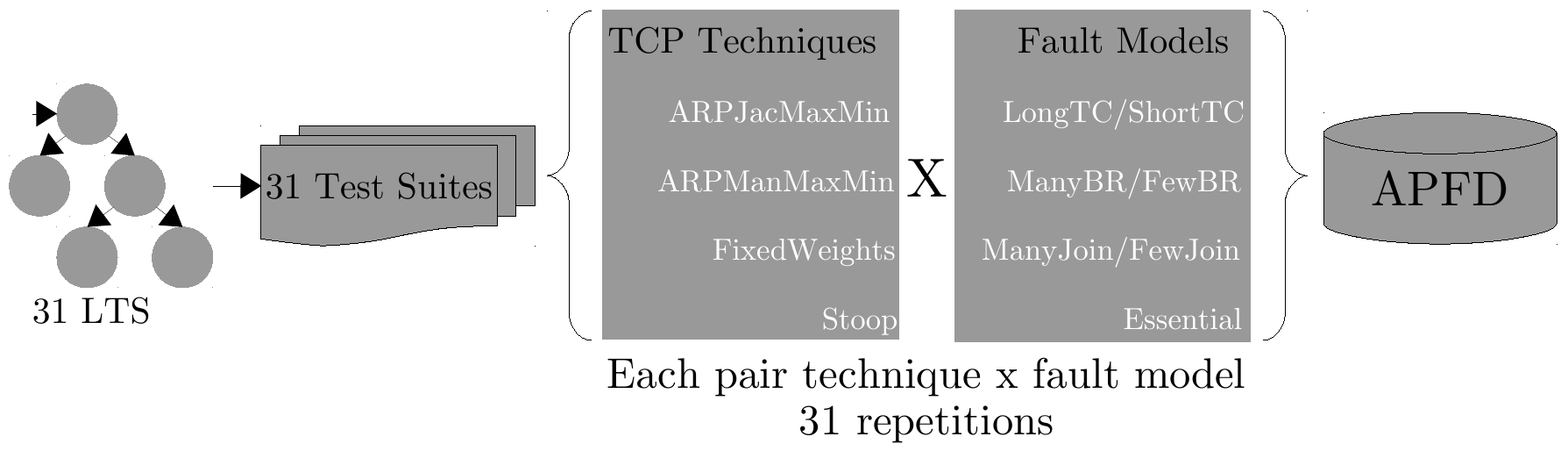}
	\caption{Overview of the original study.}
	\label{fig:OriginalStudySchema}
\end{figure}

In 
the experiment, each one of the 31 models were executed with 31 different and random failure assigned to each profile, with just one failure at once (a total of 961 executions for each technique), as depicted in Figure~\ref{fig:OriginalStudySchema}. This number of repetitions kept the design balanced and gave confidence for testing normality \cite{raj, ArcuriB2011}.

The main results were:

\begin{itemize}
	\item There was no best performer among the techniques;
	\item ARPManMaxMin was the technique less affected by the characteristics;
	\item The `'Essential test case'' characteristic did not affect significantly the techniques;
	\item There were no significant differences between the pairs of characteristics ({\bf LongTC}, {\bf ManyBR}) and ({\bf ShortTC}, {\bf FewBR});
	\item Characteristics of the test cases that fail affected the investigated techniques because techniques performed well in one scenario and bad in others.
\end{itemize}

\subsection{The Replication Study}
\label{sec:replicationStudy}

The objective of this replication is twofold: i) to reduce the threats to validity presented in the original study by using industrial artifacts and ii) to provide evidence that the results presented in both studies are not merely artifactual. To do so, we modified some aspects related to the operationalization and population, refined the research question, but kept the same
 variables, operational environment and experimenter.
Therefore, according to the classification proposed by Gómez et al. \cite{GomezJV14}, this is a changed-operationalizations/populations replication (as discussed in Section \ref{sec:replicationStudies}).

In our investigation, we intend to answer the following research questions:

\begin{itemize}
	\item \textbf{RQ1: Is there a best performer among the investigated techniques, with respect to the ability of revealing faults?} We aim at comparing the investigated techniques and either point out the one that presents the highest APFD value, or report if they are statistically similar, alternatively;
	
	\item \textbf{RQ2: Is the ability of revealing faults of the investigated techniques affected by the size of the test cases that fail?} We aim at measuring the effect sizes for the APFD 
	between the investigated techniques.
\end{itemize}

In order to represent the characteristics of the test cases that fail, the original study took several ones into consideration as discussed in Section \ref{sec:originalStudy}. However, for this replication study, we focus only on the \textbf{LongTC} and \textbf{ShortTC} characteristics. The reason is that we found out, in the original study, that the pairs (\textbf{LongTC}, \textbf{ManyBR}) and (\textbf{ShortTC}, \textbf{FewBR}) 
presented comparable results. 
Moreover, since the system models are not experimental objects of this replication, i.e., they are not automatically generated as part of the experiment, it is difficult to deal with model branches and joins. 
Furthermore, we do not consider the \textbf{Essential} characteristic in the replication since no significant differences among the techniques were observed in the original study for this characteristic. 

\subsubsection{Applications and Test Suites}
\label{sec:applications}

As mentioned in Section \ref{sec:background}, in MBT, test suites are automatically generated from models. Differently from the original study, in this replication, we consider test suites generated from models of use cases of 6 industrial applications\footnote{Developed as part of a cooperation between Ingenico do Brasil and our research lab}. Table~\ref{tab:systemsGenChar} provides general information on each application such as implementation language, lines of code, and whether they are currently in development or production. 
In summary, each application is labeled and described as follows.

\begin{itemize}
	\item S1 - an application that provides comunication between mobile devices and payment terminals;
	\item S2 - embedded system that collects biometric data for controlling student frequencies in a school;
	\item S3 - desktop application that interacts with payment terminals, sending test commands and collecting its results;
	\item S4 - system for management of general scholarly activities, including students frequency;
	\item S5 - an application that analyze failure files on payment terminals;
	\item S6 - system for management of equipment/software lending, maintenance logs, and control of bills.
\end{itemize}

\begin{table}[h]
	\centering
	\caption{General characteristics about the systems in our study.}
	\label{tab:systemsGenChar}
	\begin{tabular}{|cccc|}
		\hline
		{\bf System} & {\bf Languages}                                                            & {\bf Size (LOC)} & {\bf Under Development} \\ \hline \hline
		S1     & Java                                                                 & ~3000      & No                   \\ \hline
		S2     & C                                                                    & 3055       & No                   \\ \hline
		S3     & Java                                                                 & 13001      & Yes                  \\ \hline
		S4     & \begin{tabular}[c]{@{}c@{}}Groovy\\ Grails\end{tabular}              & 3693       & No                   \\ \hline
		S5     & \begin{tabular}[c]{@{}c@{}}Groovy\\ Java\end{tabular}                & 20713      & Yes                  \\ \hline
		S6     & \begin{tabular}[c]{@{}c@{}}Groovy\\ Javascript\\ Grails\end{tabular} & 13244      & Yes                  \\ \hline
	\end{tabular}
\end{table}
 
In the original study, the test cases were generated as one of the steps of the experiment, whereas in the replication study, they were generated, validated and executed by an actual testing team, as part of the regular testing process, and we collected the test suites used in the process to be our experiment objects. 

The development team described the behavior of each application using a controlled natural language to write use cases, as part of the team's requirements specification practice. 
From each use case description, the team automatically generated an LTS such as the one described in Section~\ref{sec:modeling}. 
%

The testing team then generated test suites from these LTS models using the same test case generation algorithm we used in the original study, which traverses loops at most twice, and uploaded them to a test case execution management tool - Testlink\footnote{http://testlink.org/}. Then, after their validation and execution by the testing team, we collected the test suites as well as the fault reports from TestLink and used them in our experiment
as the team also reported the faults using Testlink. It made our work simpler, since we could export both artifacts from the same service at the same time. The description of a fault
in these reports consists of a high level description including what unexpected behavior has manifested and a possible cause. From the reports, we collected test cases that revealed each fault.

Due to a non-disclosure agreement, we are not able to reveal further details about these applications and test suites, but we summarize some important data about the test suites
in Table~\ref{tab:systems}. For each row, we have the following data about the referred test suite:

\begin{enumerate}
	\item TS Size: amount of test cases the test suite;
	\item Mean TC Size: the arithmetic mean of the test case sizes in the referred test suite;
	\item Shortest TC: the size of the shortest test case in the referred test suite;
	\item Longest TC: the size of the longest test case in the referred test suite;
	\item \#Faults: the number of reported faults for the test suite;
	\item \#Failures: the number of test cases that fail because of the reported faults.
\end{enumerate}

\begin{table}
	\scriptsize
	\centering
	\caption{Overview about the investigated systems and related test suites.}
	\label{tab:systems}
	\begin{tabular}{|cc|c|c|c|c|c|c|}
		\hline
		&     & TS Size & Mean TC Size & Shortest TC & Longest TC & \#Faults & \#Failures \\ \hline
		\multirow{3}{*}{S1} & TS1 & 10      & 9.7          & 5           & 13         & 2      & 2        \\ \cline{2-8} 
		& TS2 & 9       & 9.5          & 5           & 11         & 2      & 2        \\ \cline{2-8} 
		& TS3 & 4       & 6.7          & 6           & 7          & 2      & 2        \\ \hline
		\multirow{3}{*}{S2} & TS1 & 10      & 10.7         & 9           & 12         & 1      & 3        \\ \cline{2-8} 
		& TS2 & 14      & 15.5         & 7           & 22         & 1      & 4        \\ \cline{2-8} 
		& TS3 & 24      & 15           & 6           & 21         & 2      & 20       \\ \hline
		\multirow{2}{*}{S3} & TS1 & 4       & 17           & 15          & 19         & 1      & 2        \\ \cline{2-8} 
		& TS2 & 5       & 22.6         & 13          & 41         & 2      & 3        \\ \hline
		\multirow{2}{*}{S4} & TS1 & 4       & 8.5          & 5           & 12         & 1      & 2        \\ \cline{2-8} 
		& TS2 & 6       & 10.8         & 5           & 17         & 2      & 2        \\ \hline
		\multirow{4}{*}{S5} & TS1 & 3       & 27           & 5           & 39         & 1      & 1        \\ \cline{2-8} 
		& TS2 & 3       & 18.3         & 17          & 19         & 2      & 2        \\ \cline{2-8} 
		& TS3 & 5       & 15           & 15          & 15         & 1      & 1        \\ \cline{2-8} 
		& TS4 & 4       & 10           & 7           & 17         & 1      & 1        \\ \hline
		\multirow{3}{*}{S6} & TS1 & 6       & 10.3         & 7           & 13         & 2      & 2        \\ \cline{2-8} 
		& TS2 & 9       & 5            & 5           & 5          & 3      & 4        \\ \cline{2-8} 
		& TS3 & 6       & 10.3         & 7           & 13         & 2      & 2        \\ \hline
	\end{tabular}
\end{table}

It is important to remark that, even though the sizes of the test suites are relatively small, the complete execution of the test suites of each application can be prohibitive, demanding the use of TCP techniques. Firstly, the test suites contains system level test cases, which means that they represent a high level usage of the application under test.  Secondly, test cases are executed manually and frequently they make use of different devices that are prepared and operated by more than one tester for a single test case execution. Finally, the setup conditions for some test cases can be rather costly, for instance specific network conditions/failures and time requirements.

Under these circumstances, the execution costs of a single test case can be quite high so that executing the complete test suite (that encompasses a single use case) may not be feasible at times, and all suites most of the times, particularly if we consider that manual test case execution:
\begin{itemize}
\item Requires the tester to be detail-oriented and inquisitive as usually execution is more than following the test case scripts.  The tester needs to closely observe the behavior of the system regarding expected results in order to define the right verdict. Identifying the expected results may not be trivial;
\item Proper documentation of failures and possible defects is required and this should be made right after each failure occurs. It might be the case that execution may be repeated a number of times so that the tester gets the right clues to describe the failure;
\item Failures may be intermittent and not easy to reproduce, demanding possible rework and redesign of the test case execution procedure.
\end{itemize}

Furthermore, regarding Table~\ref{tab:systems}, it is important to remark that faults and failures are not the same concept.
For each fault, a number of test cases fail. There are situations where for each fault, we had exactly one failure; on the other hand, there are situations where we had more than one failure for a fault. The APFD metric considers the first test case that fail for each fault. This is why we need to consider both information.

To the purpose of our investigation, we consider only the test suites that would not bias our study. To do so, we eliminate the following scenarios: test suites with just one test case and it fails or two test cases and one of them fails. These scenarios would bias our results, mainly because the study could not distinguish a good performance from a result achieved by chance.

\subsubsection{Setup}

As mentioned before, in this replication, we consider  the  \textbf{LongTC} and \textbf{ShortTC} characteristics approached in the original study.
However, in this replication, we deal with actual test cases and fault reports. Finding experimental objects that strictly meet the original 
\textbf{LongTC} and \textbf{ShortTC} could limit our scope of investigation since we would need to search for test suites and history of executions where the largest and the shortest test case would fail, respectively. 

It is reasonable to assume that approximations of the longest and shortest test cases are also good representatives of the charateristics.
Therefore, we extend the definition of the {\bf LongTC} and {\bf ShortTC} characteristics in the following way.
We establish a relation between the sizes of test cases that fail and the remainder of the test suite: i) Long test cases are those that exercise more steps of the system than the average number, possibly traversing loops - including the longest ones as in the original study, and  ii) short test cases exercise fewer execution paths than the average, usually straight and not 
traversing loops. Since we know the faults in advance, considering the set of available test suites and the test cases that fail for each one, we classify the test suites that meet each of following characteristics:

\begin{itemize}
	\item {\it LongTC}: test suites that every test case that fail is longer than the average size (From Table \ref{tab:systems}: S1-TS3, S4-TS1, S4-TS2, S5-TS1, S5-TS2, and S5-TS4);
	\item {\it ShortTC}: test suites that every test case that fail is shorter than the average size (From Table \ref{tab:systems}: S3-TS2 and S6-TS3);
	\item {\it ConstantSizeTC}: test suites where all test cases have the same size (From Table \ref{tab:systems}: S5-TS3 and S6-TS2);
	\item {\it Mixed}: The test suites that do not fit in any of the previous profiles (From Table \ref{tab:systems}: S1-TS1, S1-TS2, S2-TS1, S2-TS2, S2-TS3, S3-TS1, and S6-TS1).
\end{itemize}

In order to properly
address RQ2, 
we discarded all test suites in the {\it ConstantSizeTC} and {\it Mixed} categories since they neither resemble any characteristic investigated in the original study nor provide any significant variation of the factor. Rather,
we only consider {\it LongTC} and {\it ShortTC} categories as treatments. To measure the effect of these two treatments of the factor, we analyze the performance of the investigated techniques considering test suites that are in {\it LongTC} and {\it ShortTC} categories separately, i.e. we compare the techniques when they prioritize test suites that satisfy {\bf LongTC} conditions with their performance prioritizing test suites that satisfy {\bf ShortTC} conditions. Notice that, although we present a different way of defining {\bf ShortTC} and {\bf LongTC}, we still analyze the same variable, that may include an extended set of approximate values.

To summarize the differences and similarities between both studies, Table~\ref{tab:commonDiff} presents, side by side, the characteristics of each study. 
Aspects in common are in bold, and common values of the investigated independent variables are underlined. 

In Section~\ref{sec:techniques} we introduce the fourteen techniques (and variations) we are investigating in this replication and from now on we refer to them by their labels (bold face names defined in respective descriptions).

\begin{table}[t]
	\scriptsize
	\centering
	\caption{Side by side comparison between the original and replication studies. }
	\label{tab:commonDiff}
	\begin{tabular}{|K{1.5cm}|K{4.7cm}|K{4.6cm}|}
		\hline
		\textbf{Aspect}                        & \multicolumn{1}{c|}{\textbf{Original}}                                                                                                              & \multicolumn{1}{c|}{\textbf{Replication}}                                                                                                                                                                                               \\ \hline
		\multirow{2}{*}{\makecell{Research \\ Question}}    & \textbf{How general TCP techniques behave when test suites with certain properties fail?}                                                                    & Is there a best performer among the investigated techniques, with respect to the ability of revealing faults?                                                                                                                           \\ \cline{2-3} 
		&                                                                                                                                                     & \textbf{Is the ability of revealing faults of the investigated techniques affected by the size of the test cases that fail?}                                                                                                                     \\ \hline\hline
		\multirow{3}{*}{\makecell{Independent \\ Variables}} & \begin{tabular}[c]{@{}l@{}}\makecell{\textbf{General prioritization techniques}: \\ \underline{ARPJacMaxMin}, \underline{ARPManMaxMin}, \\ \underline{FixedWeights}, \underline{Stoop}}\end{tabular}                        & \textbf{General prioritization techniques}: Optimal, Random, \underline{ARPJacMaxMin} (renamed ARPJac), \underline{ARPManMaxMin} (renamed ARPMan) , ARTSimMaxMin, ARTSimMaxMax, \underline{FixedWeights} (renamed FW), \underline{Stoop}, PathComplexity, StringDistanceHamming, StringDistanceEuc, StringDistanceMan, StepTotal, StepAdditional \\ \cline{2-3} 
		& \begin{tabular}[c]{@{}l@{}}\textbf{Characteristics of the test cases} \\ \textbf{that fail}: \underline{LongTC}, \underline{ShortTC}, ManyBR, \\ FewBR, ManyJoin, FewJoin, Essential\end{tabular} & \begin{tabular}[c]{@{}l@{}}\textbf{Characteristics of the test cases} \\ \textbf{that fail}: \underline{LongTC}, \underline{ShortTC}\end{tabular}                                                                                                                                  \\ \cline{2-3} 
		& Number of faults defined by a fault model: fixed value equals to 1                                                                           & Actual fault reports                                                                                                                                                                                                                    \\ \hline\hline
		Dependent Variables                    & \textbf{APFD}                                                                                                                                                & \textbf{APFD}                                                                                                                                                                                                                                    \\ \hline\hline
		Experimental Objects                   & 31 synthetic LTS, similar to industrial ones, and test cases generated from them                                                                    & 17 test suites from six different industrial systems                                                                                                                                                                                    \\ \hline\hline
		Data Analysis Techniques               & \textbf{Descriptive statistics}, \textbf{visual analysis}, and \textbf{hypothesis testing}                                                                                     & \textbf{Descriptive statistics}, \textbf{visual analysis}, \textbf{hypothesis testing}, and effect size analysis                                                                                                                                                   \\ \hline
	\end{tabular}
\end{table}

\begin{figure}[t]
	\centering
	\includegraphics[width=0.7\linewidth]{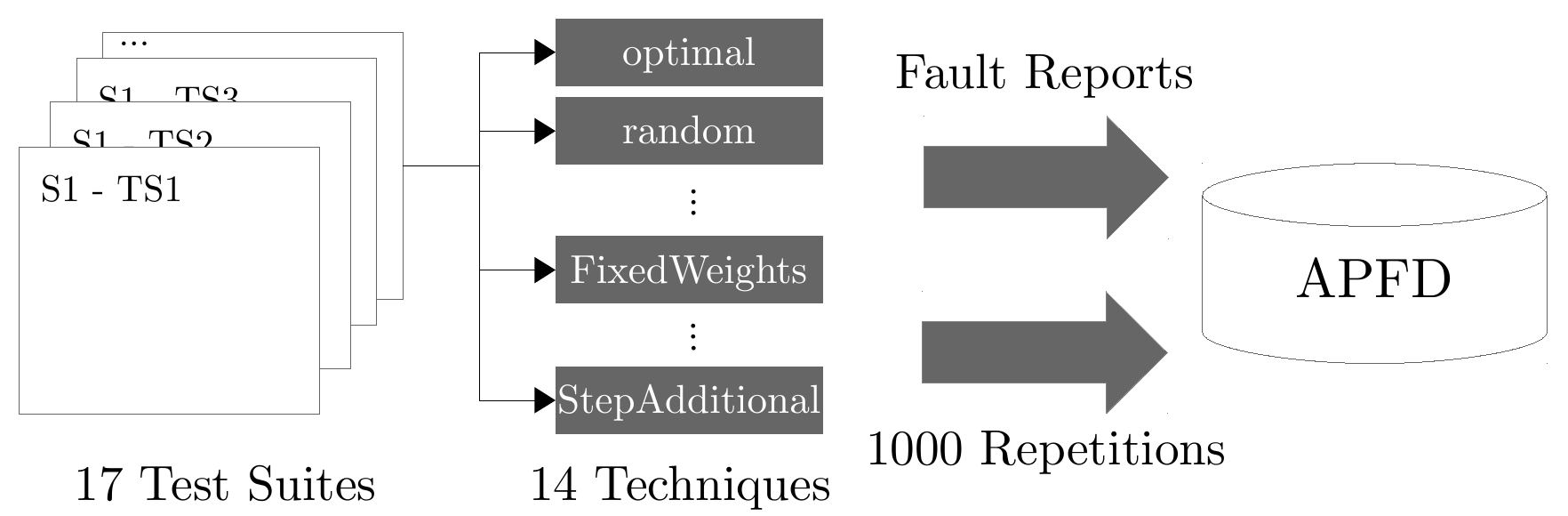}
	\caption{Overview of the replicated study.}
	\label{fig:setup}
\end{figure}

Every technique but \textbf{optimal}, present some intermediate steps that involve random choices, for example the total and additional coverage of steps solve tie situations through random choices, and adaptive random prioritization techniques create their candidate set by adding test cases randomly. Since these steps affect the proposed results, we execute every technique 1000 times with each test suite, according to the suggestion of Arcuri and Briand \cite{ArcuriB2011} for evaluation of random algorithms. From every triple \textit{(test suite, technique, repetition)} we measure the APFD. Figure~\ref{fig:setup} presents an overview of our replicated study. We executed this setup in a single machine running Ubuntu Linux 64bits, with an Intel Core i5 processor, and 6GB of RAM memory.

\subsection{Data Analysis}
\label{sec:dataAnalysis}

\begin{figure}[h]
\centering
\includegraphics[width=0.7\linewidth]{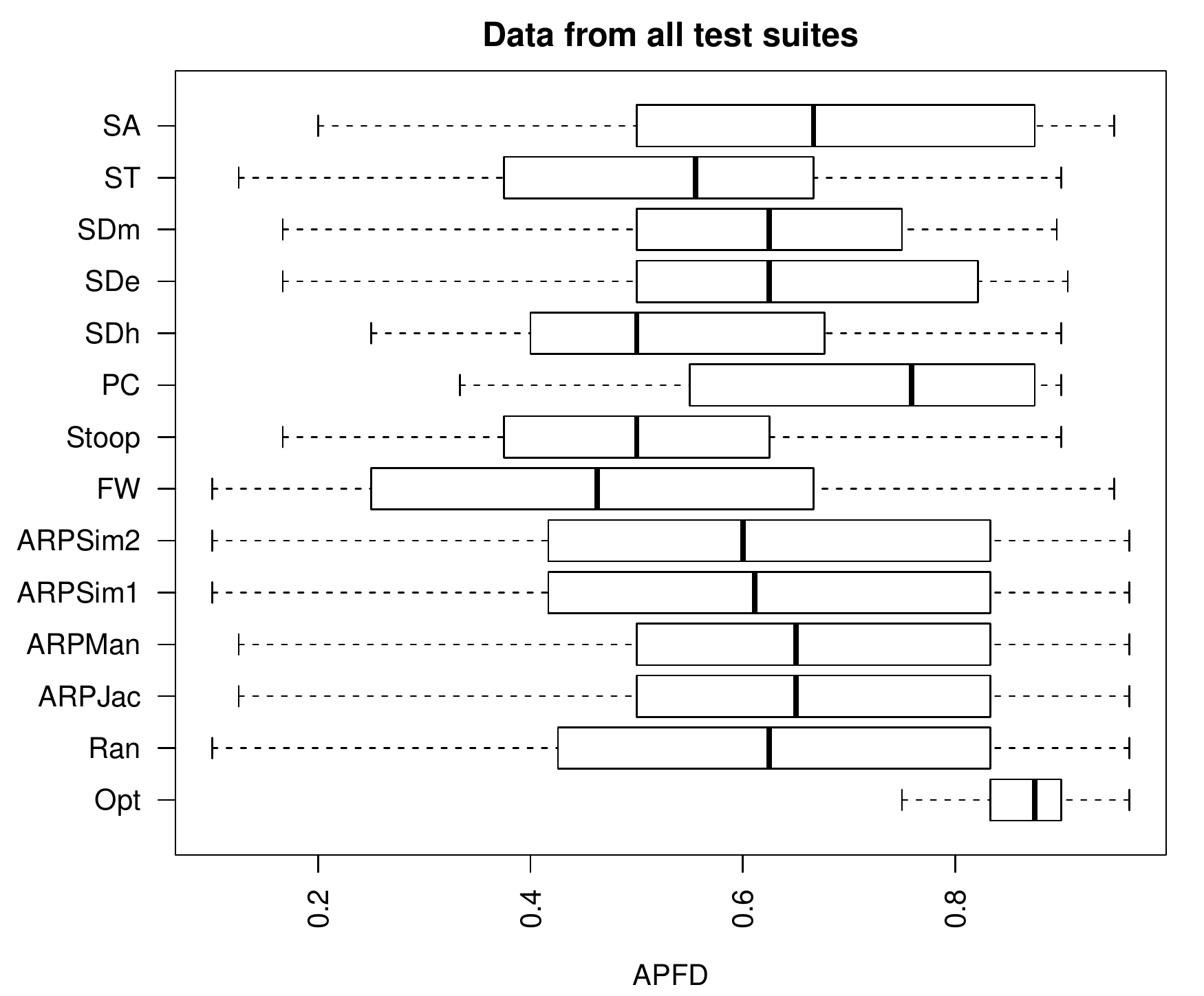}
\caption{Overall data for all techniques.}
\label{fig:overallboxplot}
\end{figure}

As a first approach to analyze the collected data\footnote{A companion web page is available through https://goo.gl/I4cPgt. It contains a brief description about this study, data files and the analysis script}, Figure~\ref{fig:overallboxplot} summarizes the overall performances of the investigated techniques. 
It can be noticed that there are differences among the investigated techniques. 
Morever, apart from \textbf{Opt}, \textbf{PC} presents the highest mean result. Nevertheless, this not yet evidence that PC is a best performer;
%
we must investigate further to address our research questions. 

When we test the hypothesis that \textit{all samples are statistically equal} through a Kruskal-Wallis, we obtain \textit{p-value}$\ = 2.2 \cdot 10^{-16}$. Therefore, even with overlaps among boxplots, techniques present different performances considering the whole set of test suites.

\begin{figure}
	\centering
      \subfloat{
		\includegraphics[width=0.48\textwidth]{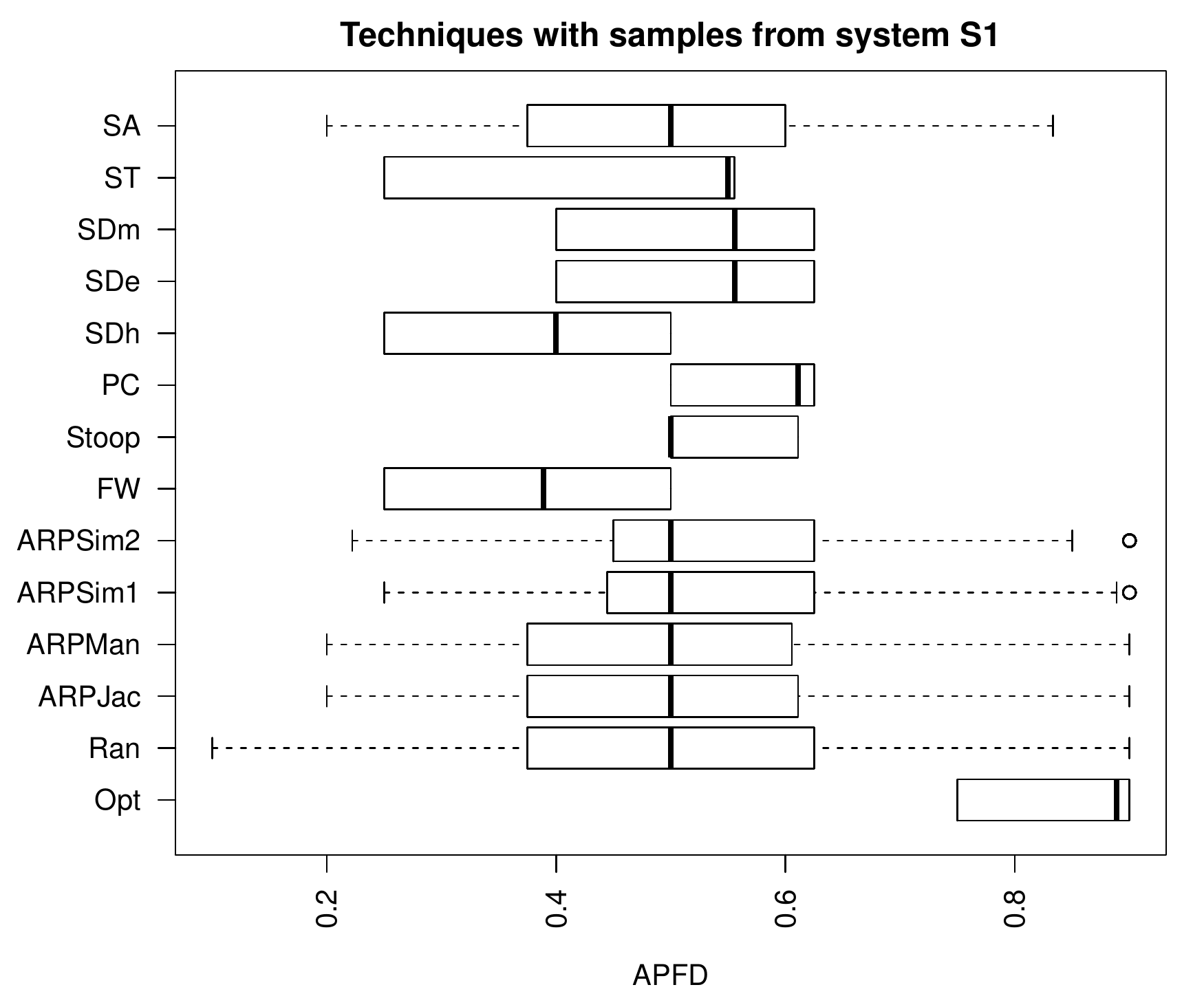}
		\label{fig:dataS1}
	}
      \subfloat{
		\includegraphics[width=0.48\textwidth]{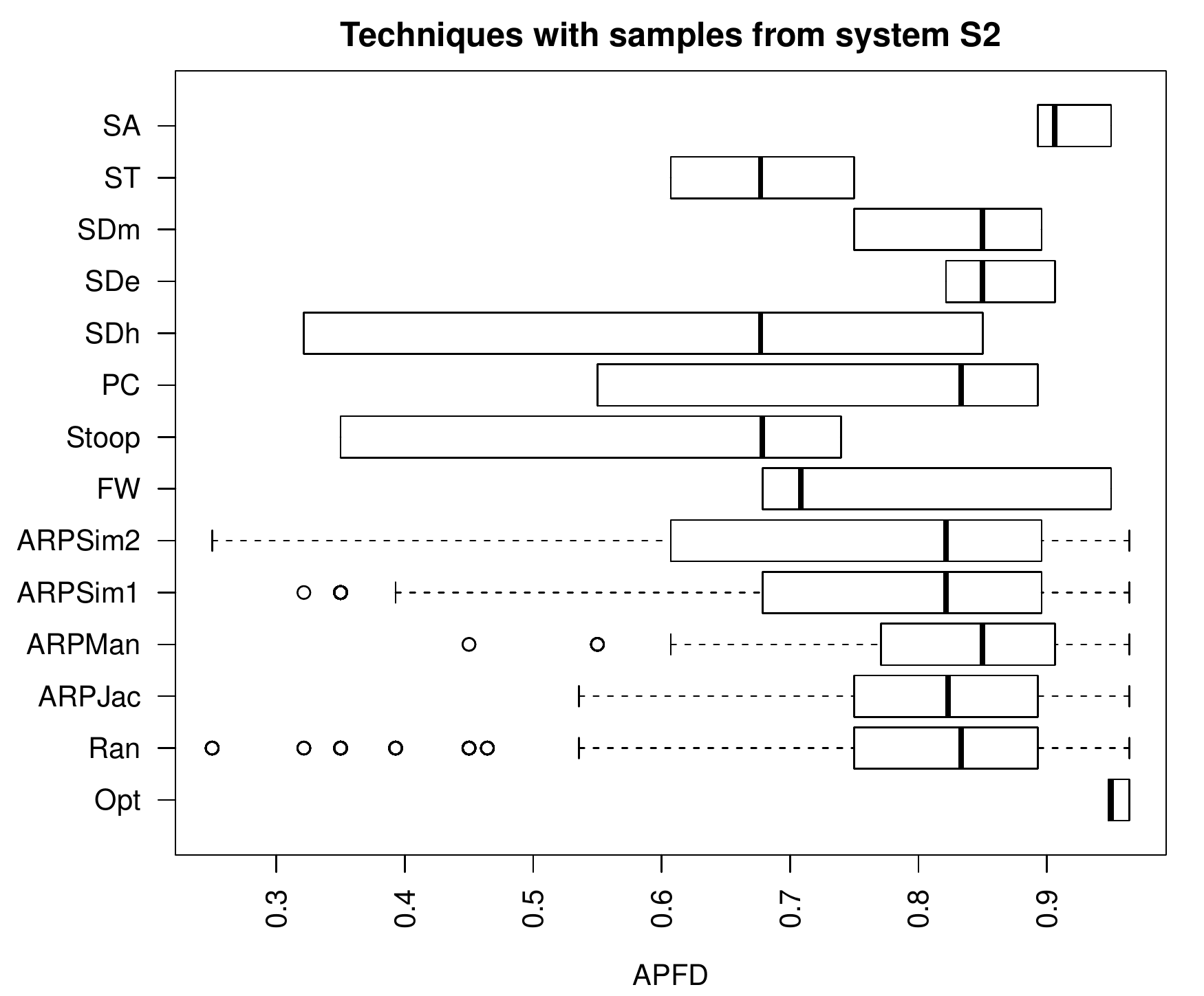}
		\label{fig:dataS2}
	}
	\\
     \subfloat{
		\includegraphics[width=0.48\columnwidth]{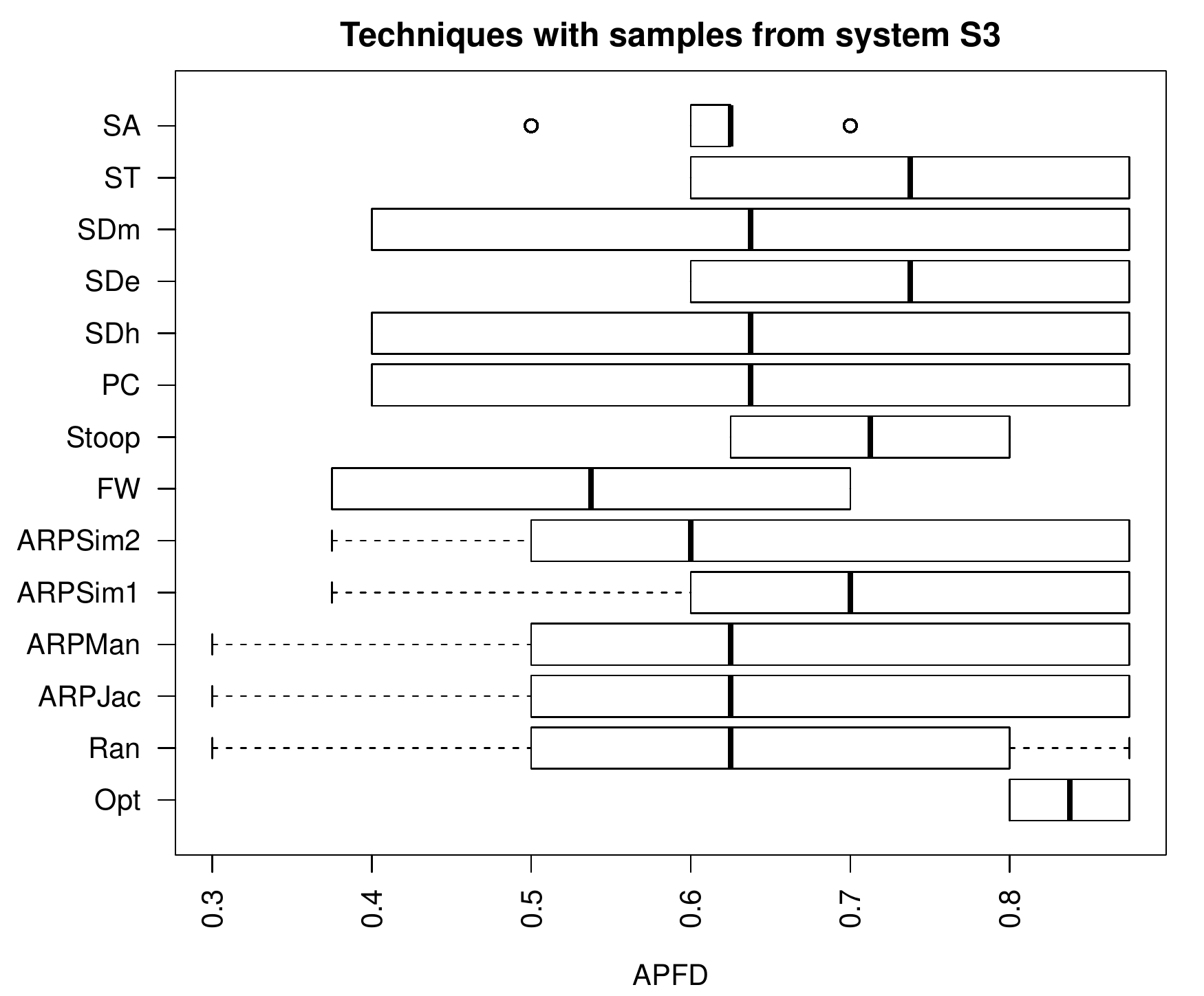}
		\label{fig:dataS3}
	}
     \subfloat{
		\includegraphics[width=0.48\columnwidth]{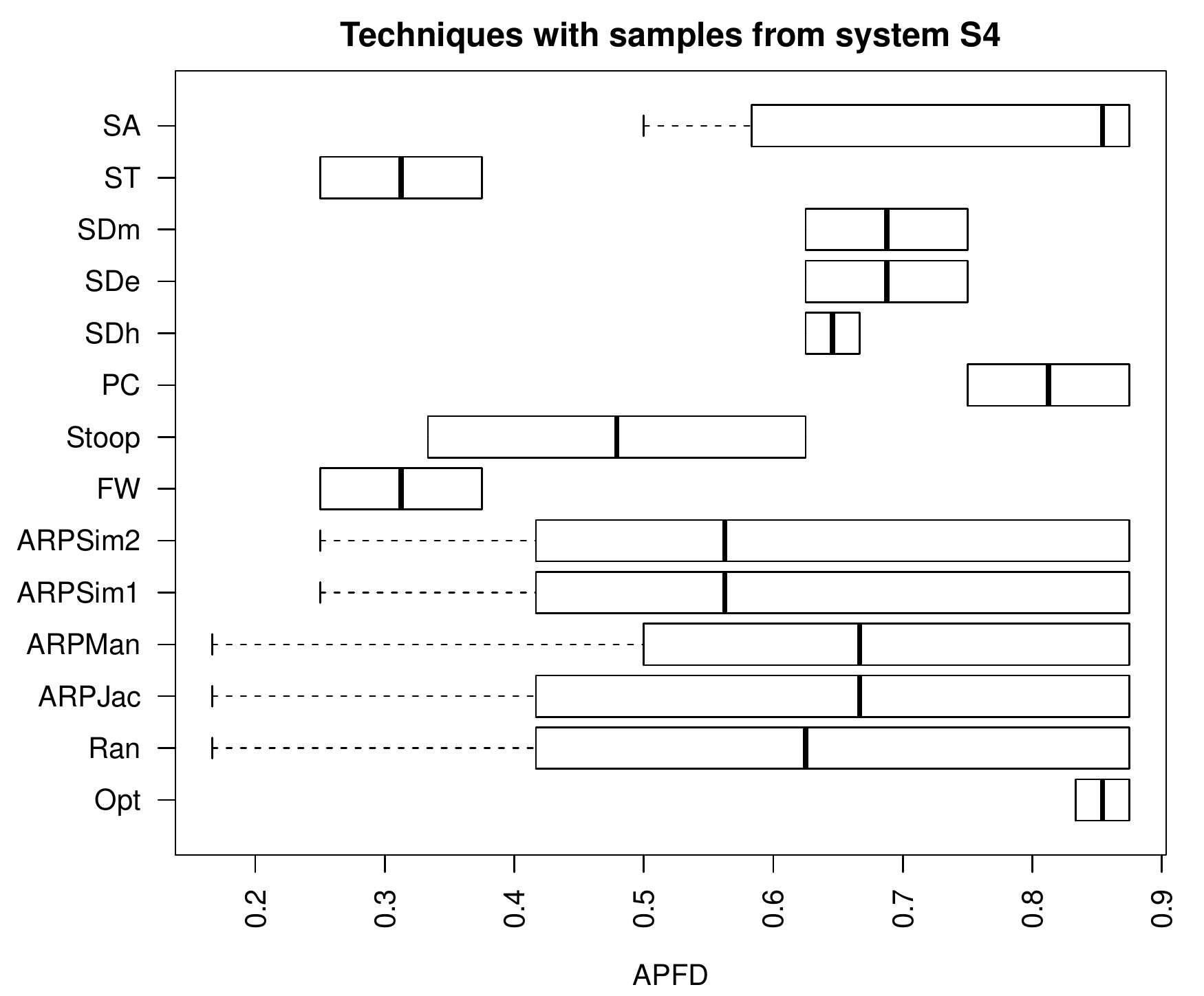}
		\label{fig:dataS4}
	}
	\\
     \subfloat{
		\includegraphics[width=0.48\columnwidth]{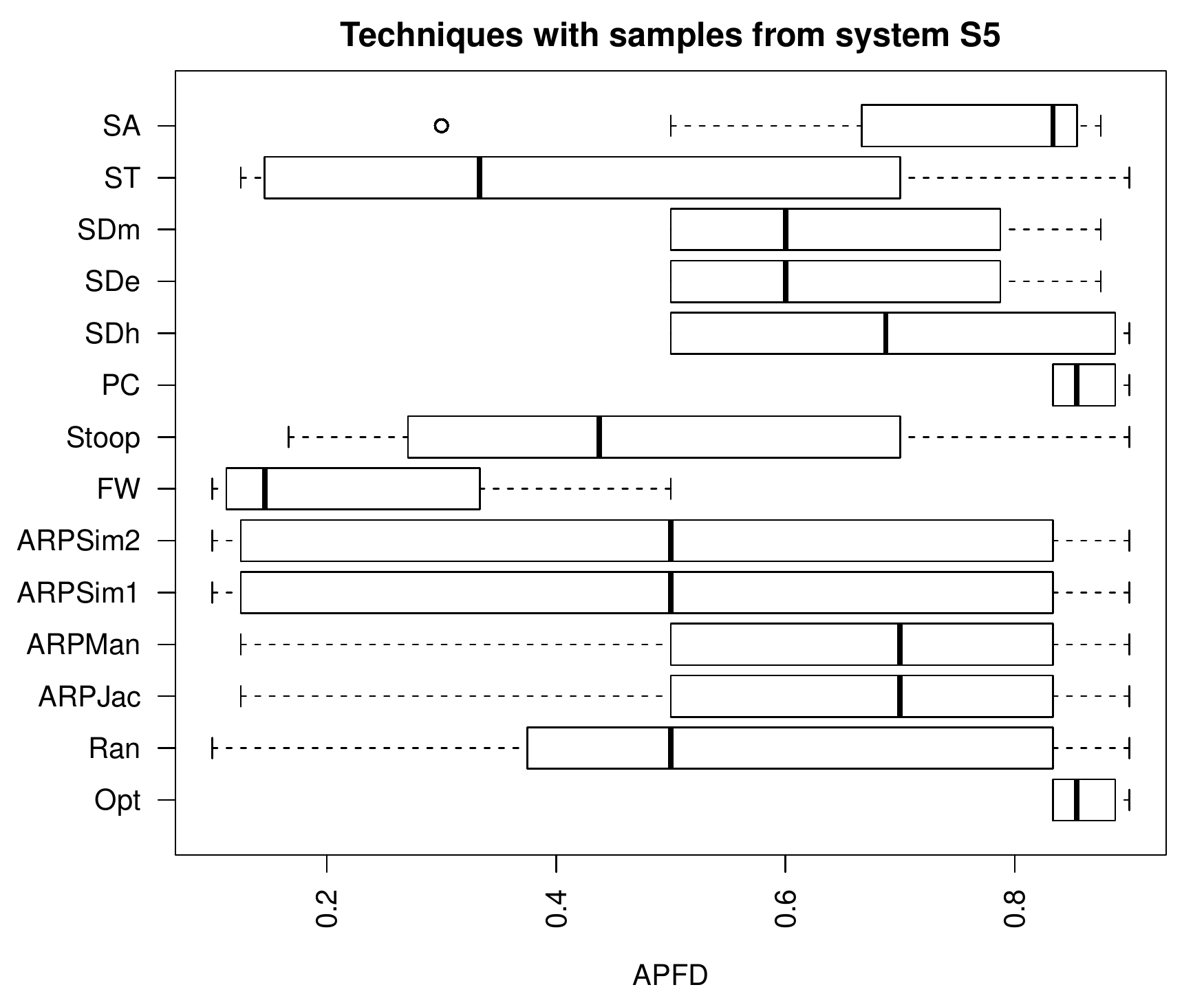}
		\label{fig:dataS5}
	}
    \subfloat{
		\includegraphics[width=0.48\columnwidth]{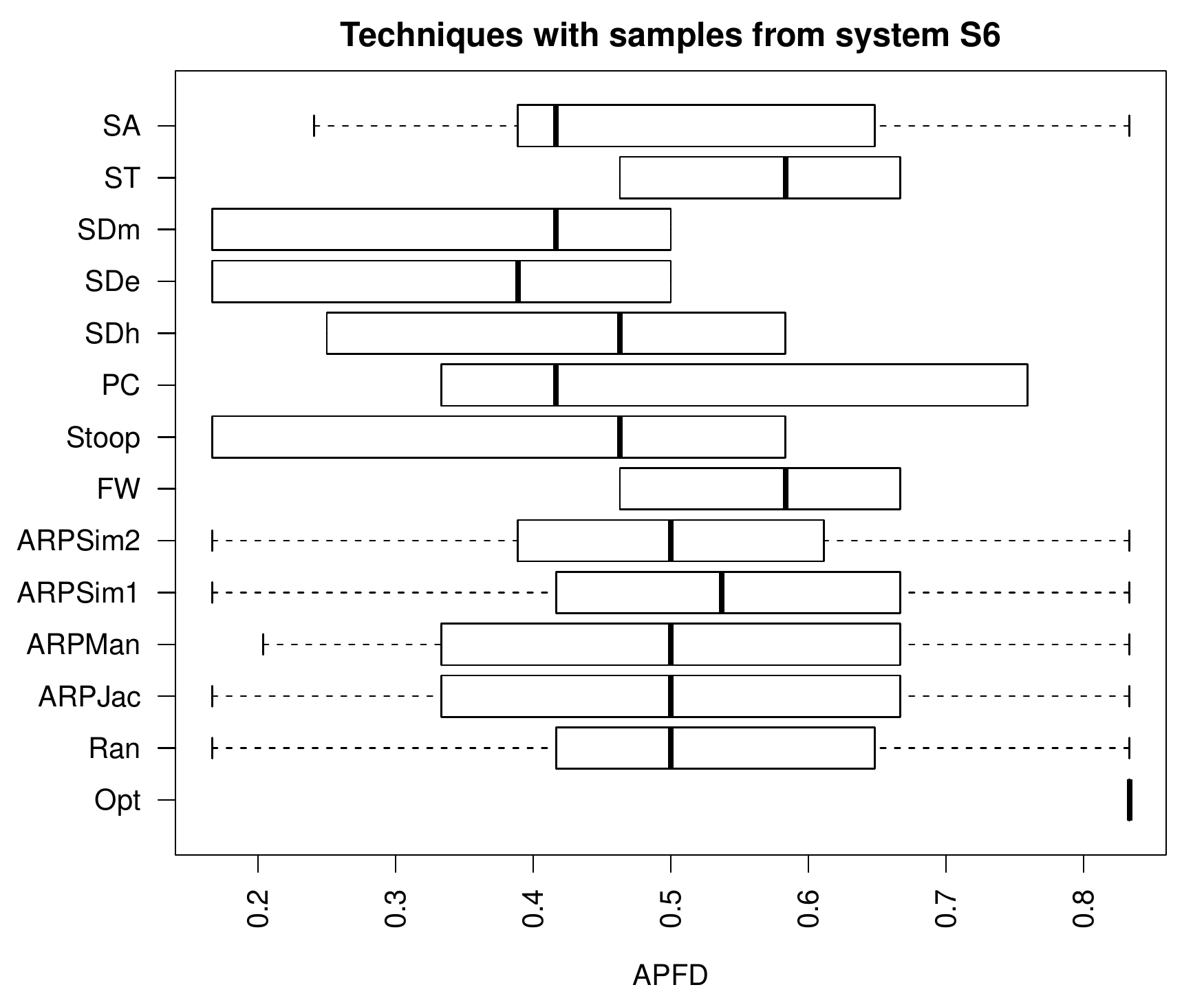}
		\label{fig:dataS6}
	}
\caption{Performance of the techniques by system}
\label{fig:persystemboxplot}
\end{figure}

Since some behaviors can be hidden by gathering data from all applications, Figure~\ref{fig:persystemboxplot} depicts the performances of the techniques grouped by 
applications. 
Notice that techniques perform differently across applications, but again \textbf{PC} appears in 3 out of 6 systems with the highest mean result (also excluding \textbf{Opt}). Analogously, we also perform Kruskal-Wallis tests on each of the six data sets and all of them result in the same p-value of the previous test, which is enough to reject all null hypotheses of equality among the samples. 

\subsubsection{RQ1}

To address this question, 
we adopt a non-parametric approach for data analysis. Because we have a high number of repetitions and there is no practical difference between both approaches,  non-parametric tests are able to test with confidence \cite{ArcuriB2011}.

A possible approach to consistently identify the best performer would be to apply pairwise hypothesis tests, e.g. Wilcoxon test, between techniques. 
 However, these tests would just suggest the best performer, without exposing how big 
 the differences between the techniques are. Thus, we need to apply effect size analysis.

To measure the effect size of the differences using a non-parametric approach, we apply the $\hat{A}_{12}$ statistic \cite{VarghaD00}. Let $s_1$ and $s_2$ be two numerical samples. The referred statistic $\hat{A}_{12}(s_1, s_2) = p$ calculate the probability $p$ that the result of a comparison between $s_1$ and $s_2$ is true. Since $p$ is a probability, $0 \le p \le 1$, we can evaluate its result by two different points of view \cite{PouldingC2010}: which is the greatest sample (Equation~\ref{eq:greatestsample}) and how big this difference is, or the effect size itself (Equation~\ref{eq:effectsize}).

\begin{equation}
	\hat{A}_{12}(s_1, s_2) =
	\begin{cases}
	p > 0.5, & \text{$\mathbf{s_1}$ is the greatest,}\\
	p < 0.5, & \text{$\mathbf{s_2}$ is the greatest,}\\
	p = 0.5, & \text{the samples are \textbf{indistinguishable}}
	\end{cases}
	\label{eq:greatestsample}
\end{equation}

\begin{equation}
	\hat{A}_{12}(s_1, s_2) =
	\begin{cases}
	\text{ if } p > 0.71 \text{ or } p < 0.29, & \text{\textbf{large} effect size};\\
	\text{ if } p > 0.64 \text{ or } p < 0.36, & \text{\textbf{medium} effect size};\\
	\text{otherwise}, & \text{\textbf{small} or negligible effect size}.
	\end{cases}
	\label{eq:effectsize}
\end{equation}

Suppose two techniques, $A$ and $B$, and our dependent variable APFD. The $\hat{A}_{12}$ statistic measures the probability that running $A$ yields a higher APFD than running $B$. For example, if we get the result $\hat{A}_{12}(A, B) = 0.67$, $A$ performs better than $B$ (from Equation \ref{eq:greatestsample}), with a medium effect size (from Equation \ref{eq:effectsize}).

Considering data from Table~\ref{tab:generalEffSize}, the major portion of the comparisons between every pair of technique results in small effect sizes. Just observing the bold-faced values, which are the medium and large effect sizes, we remark two findings: 
\begin{itemize}
	\item \textbf{FW} present consistently a bad performance, which means that just applying fixed weights based on the kinds of model elements the test suite traverses does not lead to a satisfactory ability of revealing faults; 
	
	\item \textbf{PC} is the technique that present more positive results, followed by \textbf{SA}. Therefore, the strategies that these techniques apply, such as information flow metric (refer Section~\ref{tech:pc}), and additional coverage of steps (refer Section~\ref{tech:stsa}), could be efficient in the context studied. Since $\hat{A}_{12}(PC, SA) = 0.51$, both techniques are almost statistically indistinguishable.
\end{itemize}

Therefore, we have no evidence to suggest that any technique is consistently better than the other ones, in other words:

\begin{framed}
Considering data from all test suites, there is no best performer among the investigated techniques with respect to the ability of revealing faults.
\end{framed}

\begin{table}[h]
	\tiny
	\centering
	\caption{Effect sizes of pairwise comparisons of the investigated techniques. Each cell contains the result of the comparison between the technique from the line $i$ and the one in the column $j$. The diagonal is lighter grey is when $\hat{A}_{12}(i,i)$, which always lead to effect size 0.5 and it is not relevant for our purposes. The darker grey area represents would be complementary to the presented values. Bold faced values represent medium and large effect sizes.}
	\label{tab:generalEffSize}
\begin{tabular}{r|ccccccccccccc}
        & \begin{tabular}[c]{@{}c@{}}R\\ a\\ n\end{tabular} & \begin{tabular}[c]{@{}c@{}}A\\ R\\ P\\ J\\ a\\ c\end{tabular} & \begin{tabular}[c]{@{}c@{}}A\\ R\\ P\\ M\\ a\\ n\end{tabular} & \begin{tabular}[c]{@{}c@{}}A\\ R\\ P\\ S\\ i\\ m\\ 1\end{tabular} & \begin{tabular}[c]{@{}c@{}}A\\ R\\ P\\ S\\ i\\ m\\ 2\end{tabular} & \begin{tabular}[c]{@{}c@{}}F\\ W\end{tabular} & \begin{tabular}[c]{@{}c@{}}S\\ t\\ o\\ o\\ p\end{tabular} & \begin{tabular}[c]{@{}c@{}}P\\ C\end{tabular} & \begin{tabular}[c]{@{}c@{}}S\\ D\\ h\end{tabular} & \begin{tabular}[c]{@{}c@{}}S\\ D\\ e\end{tabular} & \begin{tabular}[c]{@{}c@{}}S\\ D\\ m\end{tabular} & \begin{tabular}[c]{@{}c@{}}S\\ T\end{tabular} & \begin{tabular}[c]{@{}c@{}}S\\ A\end{tabular} \\
        \hline
Ran     & \cellcolor[HTML]{EFEFEF}                          & 0.47                                                          & 0.46                                                          & 0.51                                                              & 0.52                                                              & \textbf{0.67}                                          & 0.60                                                      & 0.37                                          & 0.54                                              & 0.46                                              & 0.48                                              & 0.59                                          & 0.40                                          \\

ARPjac  & \cellcolor[HTML]{C0C0C0}                          & \cellcolor[HTML]{EFEFEF}                                      & 0.49                                                          & 0.54                                                              & 0.55                                                              & \textbf{0.70}                                          & 0.64                                                      & 0.40                                          & 0.57                                              & 0.49                                              & 0.50                                              & 0.62                                          & 0.43                                          \\

ARPMan  & \cellcolor[HTML]{C0C0C0}                          & \cellcolor[HTML]{C0C0C0}                                      & \cellcolor[HTML]{EFEFEF}                                      & 0.54                                                              & 0.56                                                              & \textbf{0.70}                                          & \textbf{0.64}                                                      & 0.40                                          & 0.57                                              & 0.50                                              & 0.5                                               & 0.63                                          & 0.43                                          \\

ARPSim1 & \cellcolor[HTML]{C0C0C0}                          & \cellcolor[HTML]{C0C0C0}                                      & \cellcolor[HTML]{C0C0C0}                                      & \cellcolor[HTML]{EFEFEF}                                          & 0.51                                                              & \textbf{0.65}                                          & 0.58                                                      & 0.37                                          & 0.52                                              & 0.45                                              & 0.47                                              & 0.57                                          & 0.39                                          \\

ARPSim2 & \cellcolor[HTML]{C0C0C0}                          & \cellcolor[HTML]{C0C0C0}                                      & \cellcolor[HTML]{C0C0C0}                                      & \cellcolor[HTML]{C0C0C0}                                          & \cellcolor[HTML]{EFEFEF}                                          & \textbf{0.64}                                          & 0.57                                                      & 0.36                                          & 0.51                                              & 0.44                                              & 0.45                                              & 0.55                                          & 0.38                                          \\

FW      & \cellcolor[HTML]{C0C0C0}                          & \cellcolor[HTML]{C0C0C0}                                      & \cellcolor[HTML]{C0C0C0}                                      & \cellcolor[HTML]{C0C0C0}                                          & \cellcolor[HTML]{C0C0C0}                                          & \cellcolor[HTML]{EFEFEF}                      & 0.41                                                      & \textbf{0.21}                                          & 0.37                                              & \textbf{0.28}                                              & \textbf{0.29}                                              & 0.41                                          & \textbf{0.26}                                          \\

Stoop   & \cellcolor[HTML]{C0C0C0}                          & \cellcolor[HTML]{C0C0C0}                                      & \cellcolor[HTML]{C0C0C0}                                      & \cellcolor[HTML]{C0C0C0}                                          & \cellcolor[HTML]{C0C0C0}                                          & \cellcolor[HTML]{C0C0C0}                      & \cellcolor[HTML]{EFEFEF}                                  & \textbf{0.27}                                          & 0.45                                              & \textbf{0.35}                                              & 0.37                                              & 0.50                                          & \textbf{0.31}                                          \\

PC      & \cellcolor[HTML]{C0C0C0}                          & \cellcolor[HTML]{C0C0C0}                                      & \cellcolor[HTML]{C0C0C0}                                      & \cellcolor[HTML]{C0C0C0}                                          & \cellcolor[HTML]{C0C0C0}                                          & \cellcolor[HTML]{C0C0C0}                      & \cellcolor[HTML]{C0C0C0}                                  & \cellcolor[HTML]{EFEFEF}                      & \textbf{0.65}                                              & 0.59                                              & 0.61                                              & \textbf{0.71}                                          & 0.51                                          \\

SDh     & \cellcolor[HTML]{C0C0C0}                          & \cellcolor[HTML]{C0C0C0}                                      & \cellcolor[HTML]{C0C0C0}                                      & \cellcolor[HTML]{C0C0C0}                                          & \cellcolor[HTML]{C0C0C0}                                          & \cellcolor[HTML]{C0C0C0}                      & \cellcolor[HTML]{C0C0C0}                                  & \cellcolor[HTML]{C0C0C0}                      & \cellcolor[HTML]{EFEFEF}                          & 0.42                                              & 0.43                                              & 0.53                                          & 0.36                                          \\

SDe     & \cellcolor[HTML]{C0C0C0}                          & \cellcolor[HTML]{C0C0C0}                                      & \cellcolor[HTML]{C0C0C0}                                      & \cellcolor[HTML]{C0C0C0}                                          & \cellcolor[HTML]{C0C0C0}                                          & \cellcolor[HTML]{C0C0C0}                      & \cellcolor[HTML]{C0C0C0}                                  & \cellcolor[HTML]{C0C0C0}                      & \cellcolor[HTML]{C0C0C0}                          & \cellcolor[HTML]{EFEFEF}                          & 0.51                                              & 0.62                                          & 0.43                                          \\

SDm     & \cellcolor[HTML]{C0C0C0}                          & \cellcolor[HTML]{C0C0C0}                                      & \cellcolor[HTML]{C0C0C0}                                      & \cellcolor[HTML]{C0C0C0}                                          & \cellcolor[HTML]{C0C0C0}                                          & \cellcolor[HTML]{C0C0C0}                      & \cellcolor[HTML]{C0C0C0}                                  & \cellcolor[HTML]{C0C0C0}                      & \cellcolor[HTML]{C0C0C0}                          & \cellcolor[HTML]{C0C0C0}                          & \cellcolor[HTML]{EFEFEF}                          & 0.60                                          & 0.42                                          \\

ST      & \cellcolor[HTML]{C0C0C0}                          & \cellcolor[HTML]{C0C0C0}                                      & \cellcolor[HTML]{C0C0C0}                                      & \cellcolor[HTML]{C0C0C0}                                          & \cellcolor[HTML]{C0C0C0}                                          & \cellcolor[HTML]{C0C0C0}                      & \cellcolor[HTML]{C0C0C0}                                  & \cellcolor[HTML]{C0C0C0}                      & \cellcolor[HTML]{C0C0C0}                          & \cellcolor[HTML]{C0C0C0}                          & \cellcolor[HTML]{C0C0C0}                          & \cellcolor[HTML]{EFEFEF}                      & \textbf{0.32}                                          \\

SA      & \cellcolor[HTML]{C0C0C0}                          & \cellcolor[HTML]{C0C0C0}                                      & \cellcolor[HTML]{C0C0C0}                                      & \cellcolor[HTML]{C0C0C0}                                          & \cellcolor[HTML]{C0C0C0}                                          & \cellcolor[HTML]{C0C0C0}                      & \cellcolor[HTML]{C0C0C0}                                  & \cellcolor[HTML]{C0C0C0}                      & \cellcolor[HTML]{C0C0C0}                          & \cellcolor[HTML]{C0C0C0}                          & \cellcolor[HTML]{C0C0C0}                          & \cellcolor[HTML]{C0C0C0}                      & \cellcolor[HTML]{EFEFEF}                     
\end{tabular}
\end{table}

\subsubsection{RQ2}

\begin{figure}[h]
	\centering
	\includegraphics[width=1\textwidth]{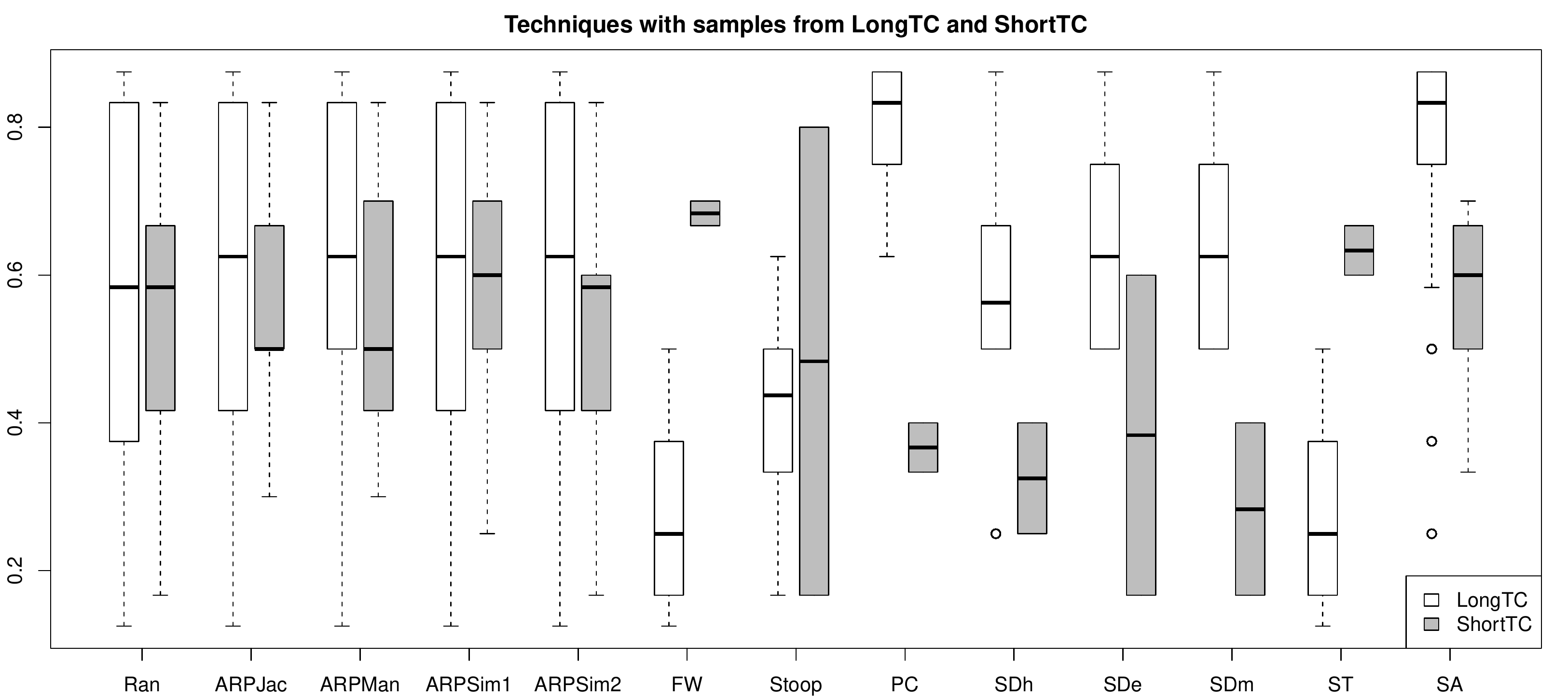}
	\caption{Boxplots of {\bf ShortTC} and {\bf LongTC} samples.}
	\label{fig:comparisonShortLong}
\end{figure}

As strategy to expose the effect of varying the size of the test cases that fail, we compare the techniques with themselves in both characteristics of the test cases that fail: \textbf{LongTC} and \textbf{ShortTC}. Figure~\ref{fig:comparisonShortLong} shows side by side the boxplots from the two considered characteristics. It can be noticed that
there are significant differences on the performance of the techniques for the different characteristics, particularly for  
\textbf{FW}, \textbf{PC}, \textbf{SDh}, and \textbf{ST}. Thus, this clear change of behavior when comparing visually both boxplots suggest that the investigated techniques are sensitive to the size of the test cases that fail.

Whereas the visual analysis just provides an initial clue about the differences among the techniques concerning the treatments, in order to quantify these influences, we also measure the effect sizes of the comparisons of every technique between the two levels of the factor. According to Table~\ref{tab:techEffSize}, based on the results of $\hat{A}_{12}$ statistic:

\begin{itemize}
	\item All random-based techniques -- 
	(\textbf{Ran}, \textbf{ARPJac}, \textbf{ARPMan}, \textbf{ARPSim1}, and \textbf{ARPSim2}) -- present small effect size. Among the adaptive random techniques, the different functions do not appear to affect significantly the results.
	
	\item Even though \textbf{Stoop} does not rely strongly on random choices, 
	it also presents small effect size. 
	
	\item The other techniques present a large effect size, in other words, they are strongly affected by the variation of the studied characteristics. 
\end{itemize}

Thus, based on the results obtained, we can conclude that:

\begin{framed}
The investigated techniques are affected by the size of the test cases that fail, but not in the same way.
\end{framed}

\begin{table}[h]
	\centering
	\caption{Effect sizes of the comparisons for each technique between {\bf ShortTC} and {\bf LongTC}}
	\label{tab:techEffSize}
	\begin{tabular}{|c|c|c||c|c|c|}
		\hline
		\textbf{Technique}	& $\mathbf{\hat{A}_{12}}$	& \textbf{Effect Size}	& \textbf{Technique}	& $\mathbf{\hat{A}_{12}}$	& \textbf{Effect Size} \\ \hline
		Ran         		& 0.4443					& Small       			& Stoop              	& 0.5833 					& Small       \\ \hline
		ARPJac      		& 0.3945					& Small       			& SDh               	& 0.0833 					& Large       \\ \hline
		ARPMan		   		& 0.374 					& Small       			& SDe				  	& 0.1666 					& Large       \\ \hline
		ARPSim1		   		& 0.4725					& Small       			& SDm				  	& 0      					& Large       \\ \hline
		ARPSim2		  		& 0.3892					& Small       			& ST		          	& 1      					& Large       \\ \hline
		FW			   		& 1							& Large       			& SA			     	& 0.1608 					& Large       \\ \hline
		PC			 		& 0							& Large       			&                    	&        					&           \\ \hline
	\end{tabular}
\end{table}

\subsection{Results and Discussion}
\label{sec:resultsAndDiscussion}

In the replicated study, 
 we investigate a set of TCP techniques and our first result is that there is no clear best performer among them. \textbf{PC} is among the best techniques but its performance is statistically indistinguishable from other techniques.
  The technique with performance more similar to \textbf{PC} 
  is \textbf{SA}. However, for some applications a particular technique performs 
  well and for others, badly, corroborating the findings in \cite{OuriquesJSERD}. The investigated techniques are sensitive to different scenarios, as proposed by the data analysis related to \textbf{RQ1}. 

\begin{figure}
	\centering
	\includegraphics[width=1\textwidth]{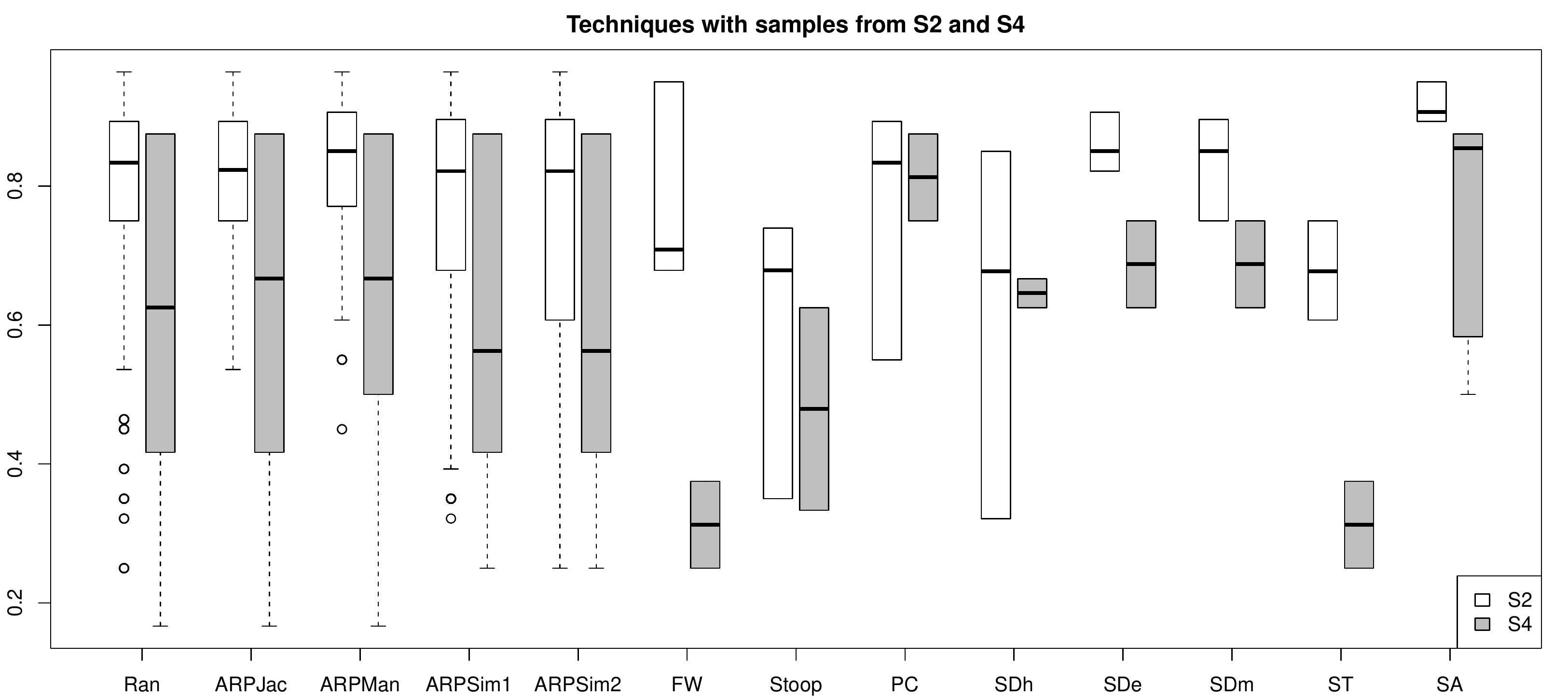}
	\caption{Boxplots of S2 and S4 samples.}
	\label{fig:comparisonS2S4}
\end{figure}

As an example of the aforementioned variation, 
consider two systems for which techniques presented different performances, say S2 and S4, from our study (see Figure~\ref{fig:comparisonS2S4} for detailed data). 
It can be noticed that 
\textbf{FW}, \textbf{SDe}, and \textbf{ST} present a good performance prioritizing test suites from S2, whereas in S4 their performances decrease to poor levels. On the other hand, \textbf{PC} and \textbf{SDh} are more accurate in S4 than in S2, as suggested by less spread boxplots in S4. The source of these variations are particularities of each test suite and system and how the investigated techniques interact with these particularities, for instance variations in the sizes of test cases
that fail.

Regarding \textbf{RQ2}, we vary the size of test cases that fail, grouping test suites by this variable, and we provide more evidence that the investigated techniques are sensitive to its variation. However, 
variation is not the same for each technique according to 
evidence provided by the analysis of the effect sizes from the comparisons of techniques in both profiles.

Techniques that present random choices in its operation are less sensitive than the other ones because 
techniques that use random choices to guide their operation make fewer assumptions about the relationship between structural aspects of the test cases and their ability of revealing faults. For instance ARP techniques have a mechanism to create candidate sets randomly, 
where the next test case to be placed in order 
is randomly selected among the ones not yet prioritized, before the distance function is applied.
Whereas we already expected this result, we are surprised by the low importance of having different distance functions on the results; maybe the selected functions, even using different properties of test cases, capture the same notion of distance between test cases. On the other hand, this result may be an indication that some level of randomness for a technique in the investigated context could help to provide a more general result.

Other technique that present an interesting behavior is \textbf{Stoop}. It seems less affected by the variation of the investigated characteristics of the test cases that fail, but still performs badly, as discussed in the RQ1 analysis. This low effect may be due to the assumption that their authors introduce; they consider important test cases that exercise common steps, i.e., steps that many other test cases also exercise. Therefore, Stoop puts them in the first positions to assure a good coverage of more important functionalities of the system under test. 
It can be noticed that this assumption does not take into account the test case sizes, but it may not lead to a good ability of revealing faults either.

The other techniques also appear to be affected by the assumptions made by their authors. \textbf{FW} gives preference to test cases that cover main scenarios of the use cases, which are usually shorter than the ones that cover alternative and exception scenarios; \textbf{ST} and \textbf{SA} sorts test cases by the number of steps, the former does not consider the feedback of the steps already covered, and the latter does; and \textbf{SDh}, \textbf{SDe}, and \textbf{SDm} put the most different test cases in the first positions and, since the distance functions are not normalized by the test case size, the techniques tend to put long test cases in the beginning.

To summarize the findings of both studies -- original and replicated --, Table~\ref{tab:commonDiffResults} 
presents a side by side comparison, and quick comments about each one of them. 
In general: 

\begin{itemize}
	\item Some minor findings from the original paper were not possible to repeat in the replication due to our modifications;
	\item Considering both studies, jaccard and manhattan functions on adaptive random technique present different levels of effect, i.e. ARPJac and ARPMan performed differently in the original study and similarly in the replication. We argue that this happened due to other characteristics of the test suites, and we 
believe that the test cases investigated in the replication present high redundancy, which affects the way that these functions calculate the distances;
	\item Although both studies do not propose a clear best performer among the investigated techniques, the ones that presented a slightly better performance in both studies, which are ARPJac in the original and PC in the replication, generally perform similarly but this comparison lies close to medium effect size;
	\item Stoop consistently presents low performances prioritizing test suites aiming to detect faults earlier. It seems that the assumption that test cases comprising steps exercised commonly by other test cases point to faults may not be valid.
\end{itemize}

\begin{table}[t]
	\centering
	\caption{Side by side comparison between the original and replication results. We represent similar results by merging columns.}
	\label{tab:commonDiffResults}
	\begin{tabular}{|K{3.5cm}|K{3.5cm}|K{3.9cm}|}
		\hline
		\textbf{Original Study}       & 
		\textbf{Replication Study}    & 
		\textbf{Comment}\\ \hline\hline
		
		\multicolumn{2}{|K{7cm}|}{Visual analysis 
		suggests differences among the techniques, considering {\bf LongTC} and {\bf ShortTC}.}  &
		In both cases, visual analysis suggests the need for deeper investigation.\\ \hline
		
		The pairs {\bf LongTC}/{\bf ManyBR} and {\bf ShortTC}/{\bf FewBR} are equivalent. &
		In the replication, we did not evaluate ManyBR and FewBR. &
		We were not able to check this relationship in the replication.\\ \hline
		
		ARPJac and ARPMan performed differently. &
		The ARP family perform similarly. &
		The test suites investigated in the replication presented a higher degree of redundancy in comparison to the ones from the original study. \\ \hline
		
		ARTJac present better overall performance. &
		PC presented better overall performance. &
		The comparison between ARPJac and PC presented a small effect size, but close to the medium threshold. On the other hand, PC seems more accurate than ARPJac. \\ \hline
		
		\multicolumn{2}{|K{7cm}|}{Stoop presented worse overall performance} &
		Stoop makes an assumption that may not lead to a good ability of revealing faults. \\ \hline
		
		\multicolumn{2}{|K{7cm}|}{There is no clear best performer} &
		Among the best performers, there was no technique that really outperformed the other ones. \\ \hline
		
		Hypothesis testing to show the effect of {\bf LongTC} and {\bf ShortTC}. &
		Effect size analysis to show the effect of {\bf LongTC} and {\bf ShortTC}. &
		Effect size analysis is a more powerful statistical tool, since it quantifies the differences instead of just indicating them. \\ \hline
		
		\multicolumn{2}{|K{7cm}|}{Varying the size of the test cases that fail ({\bf ShortTC} and {\bf LongTC}) affected the ability of revealing faults of the investigated techniques.} &
		Even with the differences between the studies, we can observe comparable outcomes. \\ \hline
	\end{tabular}
\end{table}

Based on the results obtained in the studies, 
we can argue that TCP techniques in the context investigated should not rely just on static information about the test suite (e.g. test case sizes or amount of branches) as a way of suggesting the execution order of the test cases. Instead, they should try to use other sources of information to estimate characteristics of the test cases that fail, combining them with different strategies of prioritization. We still need to evaluate the use of soft-computing methods to provide a complete overview of the techniques in the context investigated here.

From the point of view of a tester that is going to choose a TCP technique to use in a context similar to the investigated here, according the analysis of RQ1, selecting \textbf{PC} or \textbf{SA} would be equivalent. But, by its robustness across different systems and less variation among results, \textbf{PC} might be a good choice. 
On the other hand, \textbf{SA} is easier to implement as well as it can more widely applied in empirical studies.

\subsection{Threats to Validity}
\label{sec:threatsToValidity}

We evaluate the validity of our 
replicated study by discussing its threats and how we deal with them. Regarding \textbf{conclusion validity}, when analyzing the samples for {\bf ShortTC} and {\bf LongTC}, we compared samples with different sizes. To mitigate that threat, we repeated each pair technique/test suite a high number of times in order to perform non-parametric analysis with confidence. 

Concerning \textbf{construct validity}, since we considered industrial test suites and fault reports, we increased this validity in contrast to the studies performed in \cite{OuriquesJSERD}; now we have a more direct relationship between the theory and the observation. Another aspect of the construct validity is the definition of {\bf ShortTC} and {\bf LongTC}, which are the treatments of the investigated factor. Although there might be other ways of defining these characteristics, in order to keep a variability - more than one test suite in each treatment - we used the mean test case size of the test suite as a threshold to define whether the test cases that fail are short or long.

With respect to \textbf{internal validity}, we investigated only the effect of the size of test cases that fail
on the techniques, but there may be other aspects that we did not control that may affect them, for example the amount of test cases available in the sample or the proportion of test cases that fail. We report the data about the investigated test suites to provide transparency to the replicated study setup.

About \textbf{external validity}, even though we used real and industrial applications, we were not able to generalize for any other kind of application. Besides, since there is a direct relationship between the application model and the generated test suites, we were not able to generalize the results for either manually created test suites or generated through some other test case generation algorithm that follows a different heuristic. Therefore, we believe that similar conclusions hold for similar applications. Furthermore, some of these have small test suites, possibly making them not good candidates for prioritization.
This argument is even stronger when the test suites are aimed for automatic execution. Nevertheless, the seventeen test suites involved in this study were executed manually which is usually a highly demanding and costly process, specially regarding the level of the testing and the technology involved in the execution.
Therefore, even small manual test suites could benefit from prioritization, since the execution/analysis savings can  
compensate for the prioritization costs.
\section{Final Remarks}
\label{sec:finalRemarks}

This paper presented and discussed the results obtained from a replication of a previous study about techniques for general TCP in the context of MBT \cite{OuriquesJSERD}. The results presented here provide more evidence favoring the previous ones, since the conclusions point to the same direction. It is widely accepted that a number of factors may influence the performance of TCP techniques, particularly because they can be based on different aspects and strategies, including or not random choices.

In this sense, the main contribution of this paper is to investigate the influence of the size of the test cases that fail. To do so, we compare a set of 8 (families of) TCP techniques, a total of 14 techniques when considering the variations, 
using a set of 17 test suites for system testing of 6 industrial systems, developed in cooperation with Ingenico Brasil. By comparing directly the techniques, we did not have empirical evidence to suggest 
a clear best performer. In this sense, in similar contexts, a tester may opt for lower cost techniques such as the  \textbf{SA}  because it is easy to implement and uses a simple heuristic to propose a prioritized sequence of test cases. 

Since the investigated techniques are sensitive to the sizes of test cases that fail, their performances may vary depending on these characteristic. Therefore, it is important to investigate ways of reducing this variation by trying to incorporate other sources of information, yet independent of previous test executions, for instance experiences that developers have during the early stages of the system development, even before any testing. Another important aspect of the study is the lower influence of the random based techniques, which may be an indication that keeping a random aspect in a TCP technique may help to reduce the dependence of external characteristics.

As a future work, we intend to 
perform other experiments to investigate the influence of other factors on the performance of the techniques, rather than the size of the test case that fail. Moreover, we intend to 
research ways of collecting supplementary data to improve the 
effectiveness of TCP techniques, in the proposed context, making them less affected by different characteristics of test cases that fail among other factors that prove to be relevant. As consequence, we might identify which technique can be successful when data identifies the prevalence of certain characteristics.
Furthermore, we plan to repeat this analysis involving techniques with a different underlying theory (e.g. soft-computing methods), and in other contexts, for systems from different application domains.

\begin{acknowledgements}
This work was partially supported by the National Institute of Science and Technology for Software Engineering (2015), funded by CNPq/Brasil, grant 573964/2008-4. First author was also supported by CNPq grant 141215/2012-7.
\end{acknowledgements}

\bibliographystyle{spmpsci}      

\bibliography{bibliography}   

\end{document}